%% file: QCD-10-008_temp.tex
\pdfoutput=1

\documentclass[11pt,twoside,a4paper,cmspaper,final,collab]{cms-tdr}
\def\svnVersion{79860}\def\cmsCernNo{2011-049}\def\cmsCernDate{\today}\def\cmsMessage{Submitted to the Journal of High Energy Physics}

\begin{document}\cmsNoteHeader{QCD-10-008}

\hyphenation{had-ron-i-za-tion}
\hyphenation{cal-or-i-me-ter}
\hyphenation{de-vices}
\RCS$Revision: 72134 $
\RCS$HeadURL: svn+ssh://alverson@svn.cern.ch/reps/tdr2/papers/QCD-10-008/trunk/QCD-10-008.tex $
\RCS$Id: QCD-10-008.tex 72134 2011-07-27 15:15:47Z sungho $
\input{custom-definitions}
\cmsNoteHeader{QCD-10-008} % This is over-written in the CMS environment: useful as preprint no. for export versions
\title{Charged particle transverse momentum spectra \\ in pp collisions at $\sqrt{s}$ = 0.9 and 7\,\text{Te\hspace{-.08em}V}}% Force line breaks with \\

\ifthenelse{\boolean{cms@external}}{%
\author{Edward A. Wenger}\affiliation{Massachusetts Institute of Technology}
\author{Andre S. Yoon}\affiliation{Massachusetts Institute of Technology}
\author{Frank T. Ma}\affiliation{Massachusetts Institute of Technology}
}
{%

}

\date{\today}

\abstract{
The charged particle transverse momentum ($p_{\mathrm{T}}$) spectra are presented for pp collisions at $\sqrt{s}=0.9$
and 7\,\text{Te\hspace{-.08em}V}.
The data samples were collected with the CMS detector at the LHC and correspond to integrated luminosities of
231\mbox{\ensuremath{\,\mu\text{b}^\text{$-$1}}} and 2.96\mbox{\ensuremath{\,\text{pb}^\text{$-$1}}}, respectively.
Calorimeter-based
high-transverse-energy triggers are employed to enhance the statistical reach of the high-$p_{\mathrm{T}}$ measurements.
The results are compared with leading and next-to-leading order QCD and with an empirical scaling of measurements at different collision energies
using the scaling variable \mbox{$x_{\mathrm{T}}\equiv2p_{\mathrm{T}}/\sqrt{s}$} over the $p_{\mathrm{T}}$ range up to
200\ensuremath{{\,\text{Ge\hspace{-.08em}V\hspace{-0.16em}/\hspace{-0.08em}}c}}.
Using a combination of $x_{\mathrm{T}}$ scaling and direct interpolation at fixed $p_{\mathrm{T}}$,
a reference transverse momentum spectrum at $\sqrt{s}=2.76$\,\text{Te\hspace{-.08em}V} is constructed, which can be used for
studying high-$p_{\mathrm{T}}$ particle suppression in the dense QCD medium
produced in heavy-ion collisions at that centre-of-mass energy.
}

\hypersetup{%
pdfauthor={CMS Collaboration},%
pdftitle={Charged particle transverse momentum spectra in pp collisions at 0.9 and 7 TeV},%
pdfsubject={CMS 0.9 and 7 TeV charged particle spectra},%
pdfkeywords={CMS, physics}}

\maketitle %maketitle comes after all the front information has been supplied

\section{Introduction}

The charged particle transverse momentum (\pt) spectrum is an important observable for understanding the fundamental
quantum chromodynamic (QCD) interactions involved in proton-proton collisions.
While the energy dependence of the bulk of particle production with \pt\ below a few
\ensuremath{{\text{Ge\hspace{-.08em}V\hspace{-0.16em}/\hspace{-0.08em}}c}}
is typically described either empirically or with phenomenological models, the rest of the spectrum
can be well described by a convolution of parton
distribution functions, the hard-scattering cross section from perturbative calculations, and fragmentation
functions.  Such a prescription has been generally successful over a large range of lower energy pp and p$\bar{\mathrm{p}}$
collisions~\citep{Arleo:2008zd,Adare:2007dg,Adare:2008qb,Aaltonen:2009ne,PhysRevD.82.119903,:2010ir,Aamodt:2010my}.
Along with measurements of the jet production cross section
and fragmentation functions,
measurements of high-\pt\ spectra provide a test of factorised perturbative QCD (pQCD)~\cite{Yoon:2010fa} at the highest collision energy to date.

In addition to its relevance to the understanding of pQCD,
the charged particle spectrum in pp collisions will be an important reference for measurements
of high-\pt\ particle suppression in the dense QCD medium produced in heavy-ion collisions.
At the Relativistic Heavy Ion Collider (RHIC), the sizable suppression of high-\pt\ particle production,
compared to the spectrum expected from a superposition of a corresponding number of
pp collisions, was one of the first indications of strong final-state medium effects~\cite{Back:2004je,Adams:2005dq,Adcox:2004mh,Arsene:2004fa}.
A similar measurement of  nuclear modification to charged particle \pt\ spectra has been one of the first heavy-ion results at the
Large Hadron Collider (LHC)~\cite{Aamodt:2010jd}.
The reference spectrum for the PbPb collisions at $\sqrt{s_{_{\mathrm{NN}}}}=2.76$\TeV\ per nucleon can be constrained
by interpolating between the pp spectra measured at $\sqrt{s}$ = 0.9 and 7\TeV.

In this paper, the phase-space-invariant differential yield $E\,d^{3}N_{\mathrm{ch}}/dp^{3}$ is presented for primary charged
particles with energy ($E$) and momentum ($p$),
averaged over the pseudorapidity acceptance of the Compact Muon Solenoid (CMS) tracking system ($|\eta|<2.4$).
The pseudorapidity is defined as --$\ln$[tan($\theta$/2)], with $\theta$ being the polar angle of the charged particle with respect
to the counterclockwise beam direction.
The number of primary charged particles ($N_{\mathrm{ch}}$) is defined to include decay products of particles with proper lifetimes less than 1\cm.
Using the integrated luminosities calculated in Refs.~\cite{EWK-10-004,EWK-11-001} with an estimated uncertainty of 11\% and 4\%
at $\sqrt{s}=0.9$ and 7\TeV, respectively, the differential cross sections are constructed and compared to a scaling
with the variable \mbox{$\xt\equiv2\pt/\sqrt{s}$}.  Such a scaling has already been observed
for p$\bar{\mathrm{p}}$ measurements at lower collision energies~\citep{Albajar:1989an,Abe:1988yu,Aaltonen:2009ne,PhysRevD.82.119903}.
For consistency with the CDF measurements at $\sqrt{s}=0.63$, 1.8, and 1.96\TeV, the pseudorapidity
range of the \xt\ distributions has been restricted to $|\eta|<1.0$.

Finally, using the new measurements presented in this paper,
as well as previously measured pp and p$\bar{\mathrm{p}}$ cross sections, an estimate of the differential transverse momentum
cross section is constructed at the interpolated energy of $\sqrt{s}=2.76$\TeV, corresponding to the nucleon-nucleon centre-of-mass energy
of PbPb collisions recorded at the LHC.

The paper is organised as follows:
Section~\ref{sect:detector} contains a description of the CMS detector;
Section~\ref{sec:evtSel} describes the trigger and event selection;
Sections~\ref{sec:vtx} and \ref{sec:trk} detail the reconstruction and selection of primary vertices and tracks;
Section~\ref{sec:jetet} explains the characterisation of events based on the leading-jet transverse energy;
Section~\ref{sec:corr} describes the various applied corrections and systematic uncertainties;
Section~\ref{sec:results} presents the final invariant differential yields and comparisons to data and simulation;
and Section~\ref{sec:interpolation} discusses the interpolation procedures used to construct a reference spectrum at $\sqrt{s}=2.76$\TeV.

\section{The CMS Detector}
\label{sect:detector}

A detailed description of the CMS experiment can be found in Ref.~\cite{JINST}.
The central feature of the CMS apparatus is a superconducting solenoid
of 6\,m internal diameter, providing an axial magnetic field of 3.8\,T.
Immersed in the magnetic field are the pixel tracker, the silicon strip
tracker, the lead tungstate crystal electromagnetic calorimeter
(ECAL), and the brass/scintillator hadron calorimeter (HCAL).
Muons are measured in gas ionisation
detectors embedded in the steel return yoke.

The CMS experiment uses a right-handed coordinate system, with the origin at
the nominal interaction point, the $x$ axis pointing to the centre of the
LHC ring, the $y$ axis pointing up perpendicular to the plane of the LHC, and the
$z$ axis along the counterclockwise beam direction. The azimuthal angle,
$\phi$, is measured in the ($x$,\,$y$) plane.

The tracker consists of 1440 silicon pixel and 15\,148 silicon strip
detector modules and measures charged particle trajectories within the nominal
pseudorapidity range $|\eta|< 2.4$. The pixel tracker consists of three
53.3\cm-long barrel layers and two endcap disks on each side of the barrel
section. The innermost barrel layer has a radius of 4.4\cm, while for the
second and third layers the radii are 7.3\cm and 10.2\cm, respectively.
The tracker is designed to provide an impact parameter resolution of about
100\micron\ and a transverse momentum resolution of about 0.7\,\% for
1\GeVc charged particles at normal incidence ($\eta=0$)~\cite{CMSTDR1}.

The tracker was aligned as described in Ref.~\cite{TrackerAlign} using
cosmic ray data prior to the LHC commissioning. The
precision achieved for the positions of the detector modules with respect
to particle trajectories is 3--4\micron\ in the barrel for the coordinate
in the bending plane ($\phi$).

Two elements of the CMS detector monitoring system, the beam scintillator
counters (BSC) \cite{JINST, Bell} and the beam pick-up timing for the experiments
devices (BPTX)~\cite{JINST, Aumeyr}, were used to trigger the
detector readout. The BSCs are located at a distance
of 10.86\,m from the nominal interaction point (IP), one on each side, and are sensitive
in the $|\eta|$ range from 3.23 to 4.65. Each BSC is a set of $16$
scintillator tiles. The BSC elements have a time resolution of 3\,ns,
an average minimum ionising particle detection efficiency of 95.7\%, and
are designed to provide hit and coincidence rates.
The two BPTX devices, located around the beam pipe at a position of $z = \pm
175$\,m from the IP, are designed to provide precise
information on the bunch structure and timing of the incoming beam, with
better than 0.2\,ns time resolution.

The two steel/quartz-fibre forward calorimeters (HF), which extend the calorimetric coverage
beyond the barrel and endcap detectors to the $|\eta|$ region between 2.9 and 5.2,
were used for further offline selection of collision events.

The detailed Monte Carlo (MC) simulation of the CMS detector response is based
on \GEANTfour\ ~\cite{GEANT4}.
Simulated events were processed and reconstructed in the same manner as collision
data.

\section{Event Selection}
\label{sec:evtSel}

This analysis uses data samples collected from 0.9 and 7\TeV pp collisions in the first months of the 2010 LHC running,
corresponding to integrated luminosities of $(231 \pm 25)$\microbinv\ and $(2.96 \pm 0.12)$\pbinv,
respectively~\cite{EWK-10-004,EWK-11-001}.
This section gives a brief description of the requirements imposed to select good events for this analysis.
A more detailed description of the CMS trigger selections can be found in Ref.~\cite{Khachatryan:2010xs}.

First, a minimum bias trigger was used to select events with a signal in any of the BSC tiles,
coincident with a signal from either of the two BPTX detectors, indicating
the presence of at least one proton bunch crossing the interaction point.  From this sample, collision events were
selected offline by requiring a coincidence of BPTX signals, indicating the presence of both beams.

To select preferentially non-single-diffractive (NSD) events, at least one forward calorimeter (HF) tower
with energy deposition $E>3$\GeV in each of the forward and backward hemispheres was required.  Events with beam-halo muons
crossing the detector were identified and rejected based on the time difference between BSC hits on either side of
the interaction point.  Beam-induced background events, producing anomalous numbers of low-quality tracks, were rejected by
requiring that at least 25\% of the charged particles reconstructed in the pixel--silicon tracking system satisfied the \textit{highPurity}
criterion.  This criterion, described in Ref.~\cite{TRK-10-001}, consists of numerous selections on the properties of the tracks,
including the normalised $\chi^2$, the compatibility with the beamline and primary vertices, the number of hit layers,
the number of `3D' layers, and the number of lost layers.
The selection on the fraction of \textit{highPurity} tracks was only applied to events with more than 10 tracks, providing a clean separation
between real pp collisions and beam backgrounds.  The remaining non-collision event fraction,
determined by applying the same selections to events where only a single beam was crossing the interaction point,
is estimated to be less than 2 x $10^{-5}$.
Events were required to have at least one primary vertex, reconstructed according to the description in the following section
from triplets of pixel hits.
A further requirement, namely at least one vertex found from fully reconstructed tracks (see next section for details)
with number of degrees of freedom
($Ndof$) greater than four, was imposed to improve the robustness against triggered events containing
multiple pp collisions, i.e., ``event pileup''.  The loss in event selection efficiency
from the fully-reconstructed-track vertex compared to the pixel vertex alone was determined entirely from data, based on a subset
of early runs with negligible event pileup.
The percentage of events remaining after each selection step is presented in Table~\ref{tab:evtSel}.

For a large part of the 7\TeV data collection, the minimum bias trigger paths had to be prescaled by large factors because of
the increasing instantaneous luminosity of the LHC.  In order to maximise the \pt\ reach of the charged particle transverse
momentum measurement at this centre-of-mass energy, two high-level trigger (HLT) paths were used that selected events with
minimum uncorrected transverse jet energies (\et) of 15 and 50\GeV, based only on information from the
calorimeters.   While the higher threshold path was not prescaled during the 7\TeV data-taking period corresponding to
the 2.96\pbinv\ used in this analysis, the lower threshold path had to be prescaled for a significant fraction of this sample.
The 0.9\TeV\ data sample consists of 6.8 million minimum bias triggered events, while the 7\TeV sample is composed
of 18.7 million minimum bias events, and 1.4 (5.6) million events selected with the HLT minimum-\et\ values of 15 (50)\GeV.

\begin{figure}[t]
	\centering
	\subfigure{
	  \includegraphics[width=0.45\textwidth]{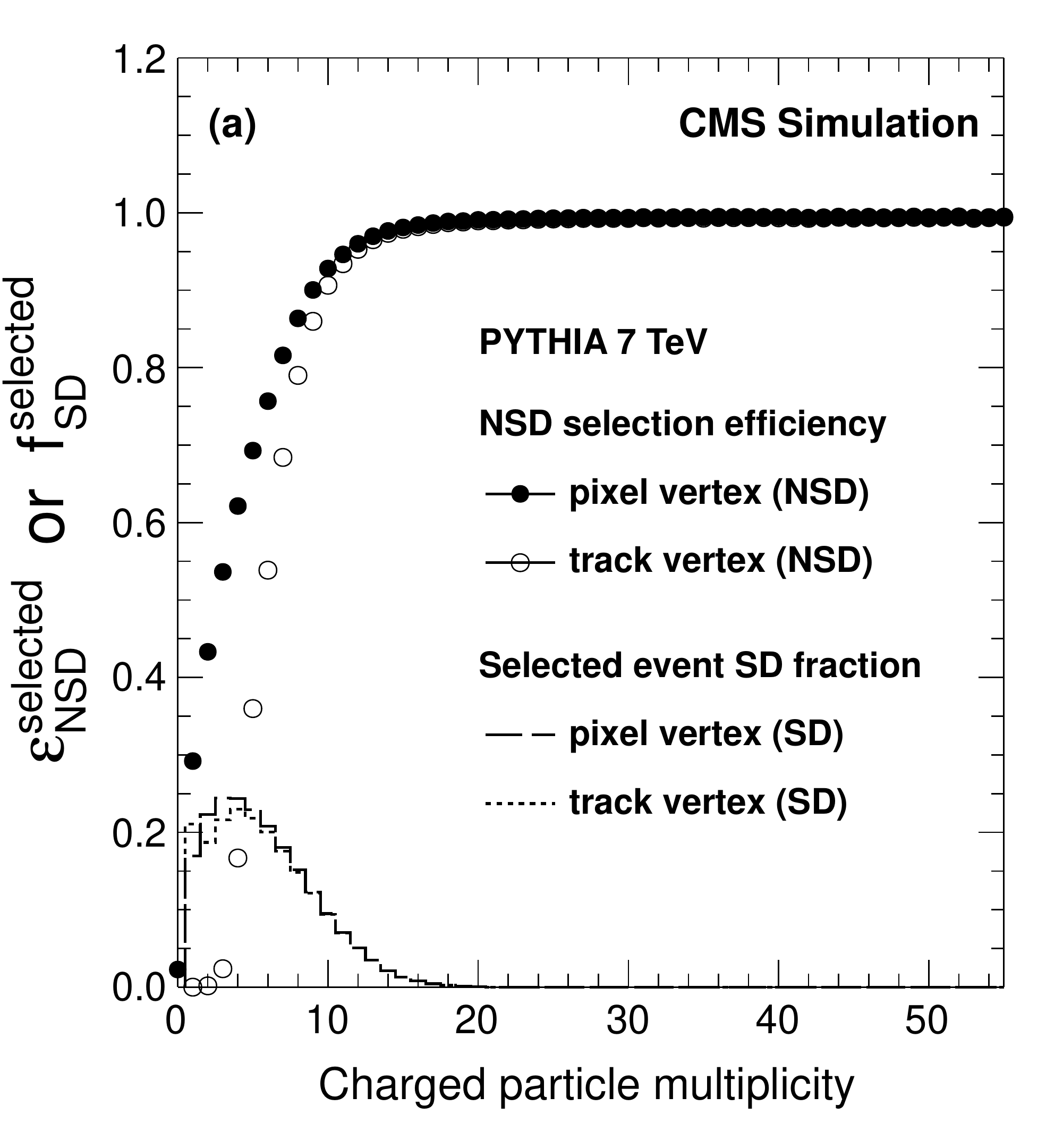}
	  \label{fig:evtSelEff}}
	\subfigure{
	  \includegraphics[width=0.45\textwidth]{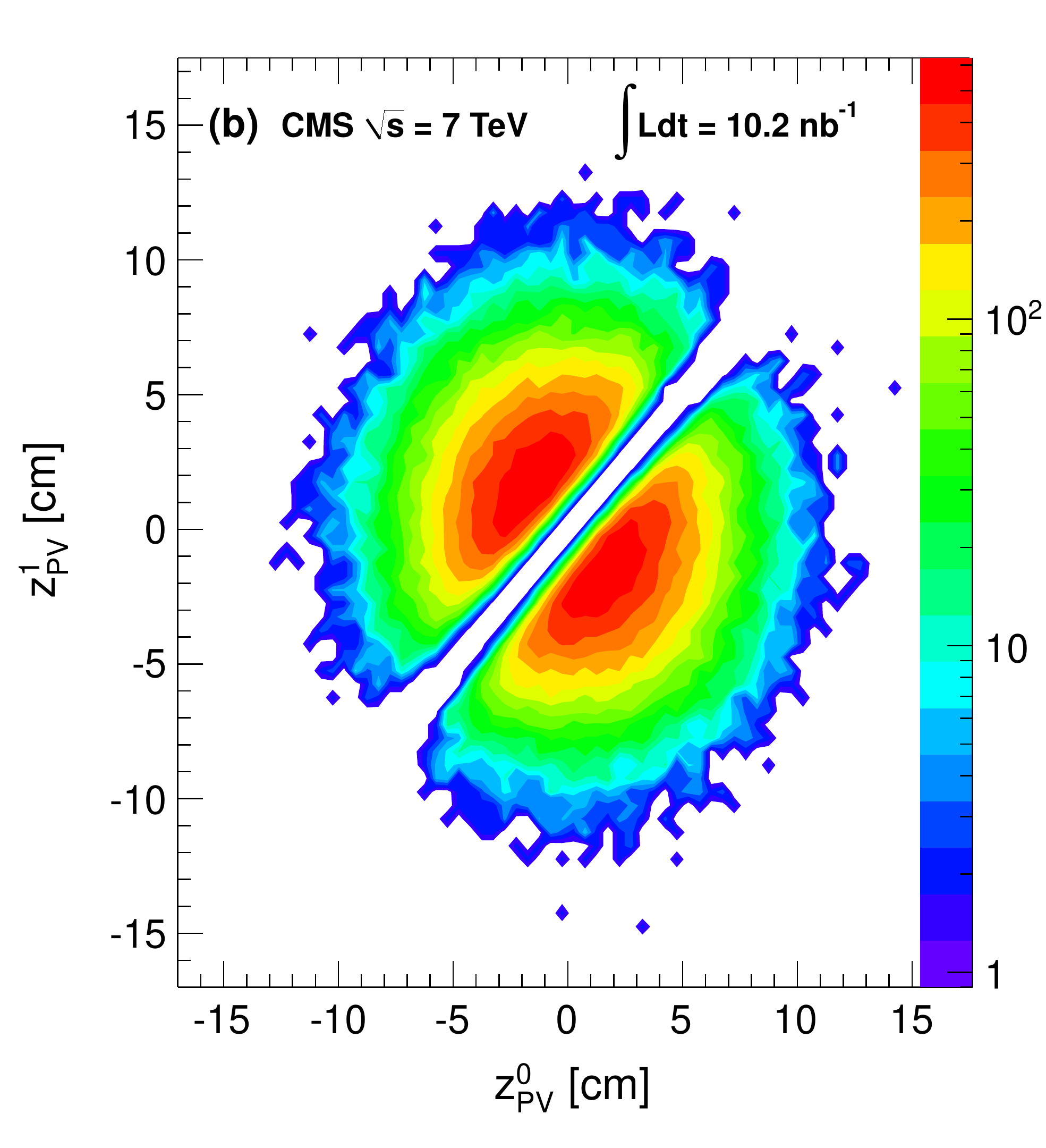}
	  \label{fig:multPrimVtxZ}}
	\caption{
	(a) The efficiency ($\varepsilon_{\mathrm{NSD}}^{\mathrm{selected}}$ in Eq.~(\ref{eqn:evtWeight})) for selecting non-single-diffractive (NSD)
	events as a function of the multiplicity of reconstructed charged particles in the tracker acceptance (\mbox{$|\eta|<2.4$})
	after applying the full event selection described in the text,
	including a single pixel-track vertex (filled circles) and additionally requiring a fully-reconstructed-track vertex
	with $Ndof>4$ (open circles)
	as described in Section~\ref{sec:vtx}.
	Also, the remaining single-diffractive (SD) fraction ($f^{\mathrm{selected}}_{\mathrm{SD}}$ in Eq.~(\ref{eqn:evtWeight})) as a function of charged
	particle multiplicity for the same selections (solid and dashed lines).
    (b) Correlation between the $z$ positions, $z^0_{\mathrm{PV}}$ and $z^1_{\mathrm{PV}}$, of the two vertices with the most associated tracks
    for measured events with more than one fully-reconstructed-track vertex satisfying the quality selections.
	}
	\vspace{4mm}
	\label{fig:evtSel}
\end{figure}

The selection efficiency for NSD events was determined based on simulated events from the \textsc{pythia}~\cite{Sjostrand:2006za}
event generator (version 6.420, tune D6T~\cite{Bartalini:2009xx}) that were subsequently passed through
a Monte Carlo simulation of the CMS detector response.
The resulting event selection efficiency as a function of the multiplicity of reconstructed charged particles
is shown for 7\TeV collisions in Fig.~\ref{fig:evtSelEff}.
The corresponding event selection efficiency is calculated by the same technique for the 0.9\TeV data (not shown).
Based on events simulated with \textsc{phojet}~\cite{Bopp:1998rc,Engel:1995sb} and \textsc{pythia}, the remaining fraction
of single-diffractive (SD) events in the selected sample was estimated to be ($5 \pm 1$)\% and ($6 \pm 1$)\%
for the 0.9 and 7\TeV data, respectively.

\begin{table}[tb]
\caption{Summary of event selection steps applied to the 0.9 and 7\TeV collision data sets and
the percentage of events from the original minimum bias samples that remain after each step. }
\centering
\begin{tabular}{ l c c c }
\\ \hline \hline
Collision energy & 0.9\TeV & 7\TeV \\
\hline
Selection & \multicolumn{2}{c}{Percentage passing each selection cut} \\
\hline \hline
One BSC + one BPTX & 100.0 & 100.0 \\
BPTX coincidence & 94.49 & 90.05 \\
Beam halo rejection & 94.08 & 89.83 \\
HF coincidence & 73.27 & 83.32 \\
Beam background rejection & 73.26 & 83.32 \\
Valid pixel-track vertex & 70.14 & 82.48 \\
Quality full-track vertex & 64.04 & 77.35 \\
\hline \hline
\vspace{2mm}
\end{tabular}
\label{tab:evtSel}
\end{table}

\section{Primary Vertex}
\label{sec:vtx}

In this analysis, two separate algorithms are employed to determine the primary vertex position.
The first is a highly efficient algorithm based on pixel triplet tracks that requires a minimum of just a single track
consistent with the beam-spot position.
The position of the beam-spot, taken as the centre of the region where the LHC beams collide,
is calculated for each LHC fill based
on the average over many events of the three-dimensional fitted vertex positions~\cite{TRK-10-001}.
The second vertex-finding algorithm, based on fully reconstructed tracks with hits also in the silicon strip tracker,
is less efficient in selecting low-multiplicity events, but more robust in discriminating against event pileup.
Since pileup is significant over the majority of the analysed data sample, only the fully-reconstructed-track vertex is used
to construct the raw charged particle momentum spectra.
The raw spectra are subsequently corrected for the fraction of events with fewer than four tracks (and the fraction of tracks
in such low-multiplicity events), based on a subset of the event sample selected with the more efficient pixel-track vertex
requirement during collision runs with negligible event pileup.

To determine the $z$ position of the pixel vertex in each event, tracks consisting of three pixel hits
are constructed with a minimum \pt\ of 75\MeVc\ from a region within a transverse distance of 0.2\cm\ from the beam axis.
The $x$ and $y$ positions of the pixel vertex are taken from the transverse position of the beam axis.
Fitted tracks are selected based on the requirement that the transverse impact parameter is less than three times the
quadratic sum of the transverse errors on the track impact parameter and the beam axis position.
The selected tracks are then passed to an agglomerative algorithm~\cite{Sikler:2009nx},
which iteratively clusters the tracks into vertex-candidates.  The procedure is halted when the distance between
nearest clusters, normalised by their respective position uncertainties, reaches 12.  Only vertices consisting of at least
two tracks are kept, except when the event contains a single reconstructed track, which occurs in 1.67\% (0.99\%) 
of the events at $\sqrt{s}=0.9$ (7)\TeV. 
In the case of multiple vertex-candidates, only the vertex with the most associated tracks is kept.
While this occurs in as many as 20\% of events, the rejected vertex typically has very few
associated tracks and is highly correlated in $z$ position to the vertex with the most associated tracks.
These characteristics imply that the rejected vertices are not from event pileup, but rather from tracks
in the tails of the impact parameter distribution that are not agglomerated into the primary vertex.

The fully-reconstructed-track vertex algorithm begins from a set of tracks selected according to their transverse impact
parameter to the beam-spot ($<2\cm$), number of hits ($>6$), and normalised $\chi^2$ ($<20$).  These tracks are
passed to an adaptive vertex fitter, in which tracks are assigned a weight between 0 and 1 according to their compatibility
with the common vertex~\cite{TRK-10-001}.  Quality vertices are further required to have more than four degrees
of freedom ($Ndof$), corresponding to at least four tracks with weights of approximately one.
For events with multiple reconstructed vertices passing the quality selection, the correlation between the $z$ positions
of the two vertices with the most associated tracks is shown in Fig.~\ref{fig:multPrimVtxZ}.  Other than the diagonal
region without multiple vertices, expected from the algorithmic parameter of at least a 1\cm\ separation,
the uncorrelated positions of the two vertices are indicative of random event pileup.

The event pileup rate is estimated from the fraction of events with multiple reconstructed vertices, after correcting
for vertices that are not found because of their proximity.
The beam conditions varied over the analysed minimum bias data samples, such that the corrected fraction of pileup events
is in the range (0.4--7.5)\%.  The uncertainty on the event pileup fraction, determined from the largest correction to the
multiple-vertex fraction, is a constant factor of 0.2\% and 1.2\% for the 0.9 and 7\TeV data, respectively.

\section{Track Selection}
\label{sec:trk}

This analysis uses tracks from the standard CMS reconstruction algorithm, which consists of multiple iterations of a
combinatorial track finder based on various seeding layer patterns~\cite{Adam:2006ki}.  After each iteration,
hits belonging unambiguously to tracks in the previous step are removed from consideration for subsequent steps.

In order to minimise the contribution from misidentified tracks and tracks with poor momentum resolution, a number of quality
selections are applied.  These include the \textit{highPurity} selection mentioned in Section~\ref{sec:evtSel}, the requirement
of at least five hits on the track, the normalized $\chi^{2}$ per degree of freedom divided by the number
of tracker layers used in the fit less than a maximum value which varies from 0.48 and 0.07 depending 
on $\eta$ and \pt, and a relative momentum uncertainty of less than 20\%.  Furthermore, to reject non-primary
tracks (i.e., the products of weak decays and secondary interactions with detector material), only the pixel-seeded tracking
iterations are used, and selections are placed on the impact parameter of the tracks with respect to the primary vertex position.
Specifically, the transverse and longitudinal impact parameters are required to be less than 0.2\cm\ and also less than 3
times the sum in quadrature of the uncertainties on the impact parameter and the corresponding vertex position.
In the case of multiple quality reconstructed vertices in the minimum bias event samples, tracks that pass the impact
parameter selections with respect to any vertex are used in the analysis.  The number of events, by which the track \pt\
distribution is normalised, is then scaled by a factor to account for the event pileup fraction.  In contrast, for the jet-triggered
samples, tracks are selected based on the impact parameter with respect to the single vertex responsible for the trigger.
The primary vertex of the hard-scattering process is identified as the vertex with the largest value of $\sum{\pt^2}$ for the
associated fitted tracks.

\begin{figure}[t]
	\centering
	\subfigure{
        \includegraphics[width=0.45\textwidth]{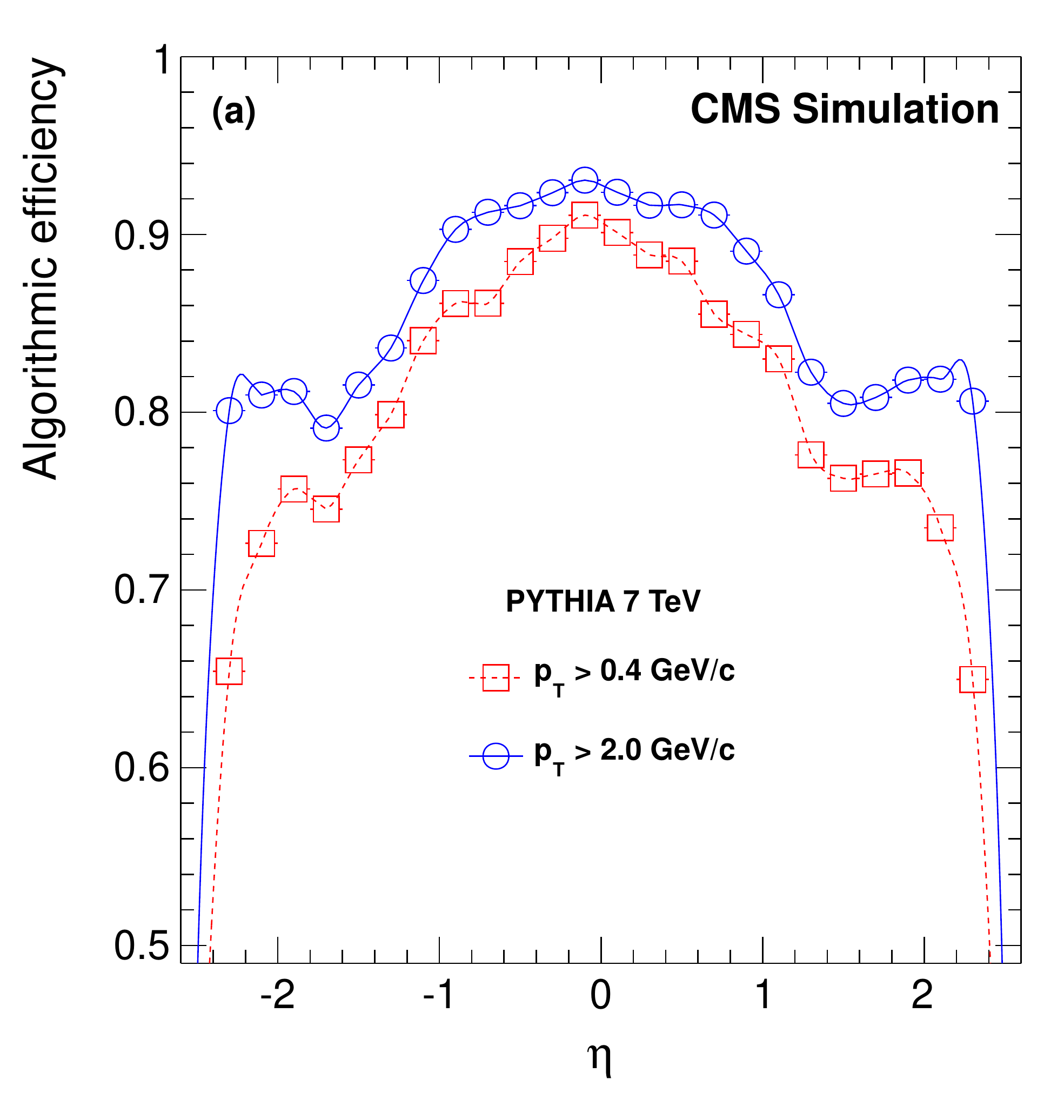}
	\label{fig:trkEff}}
	\subfigure{
	\includegraphics[width=0.45\textwidth]{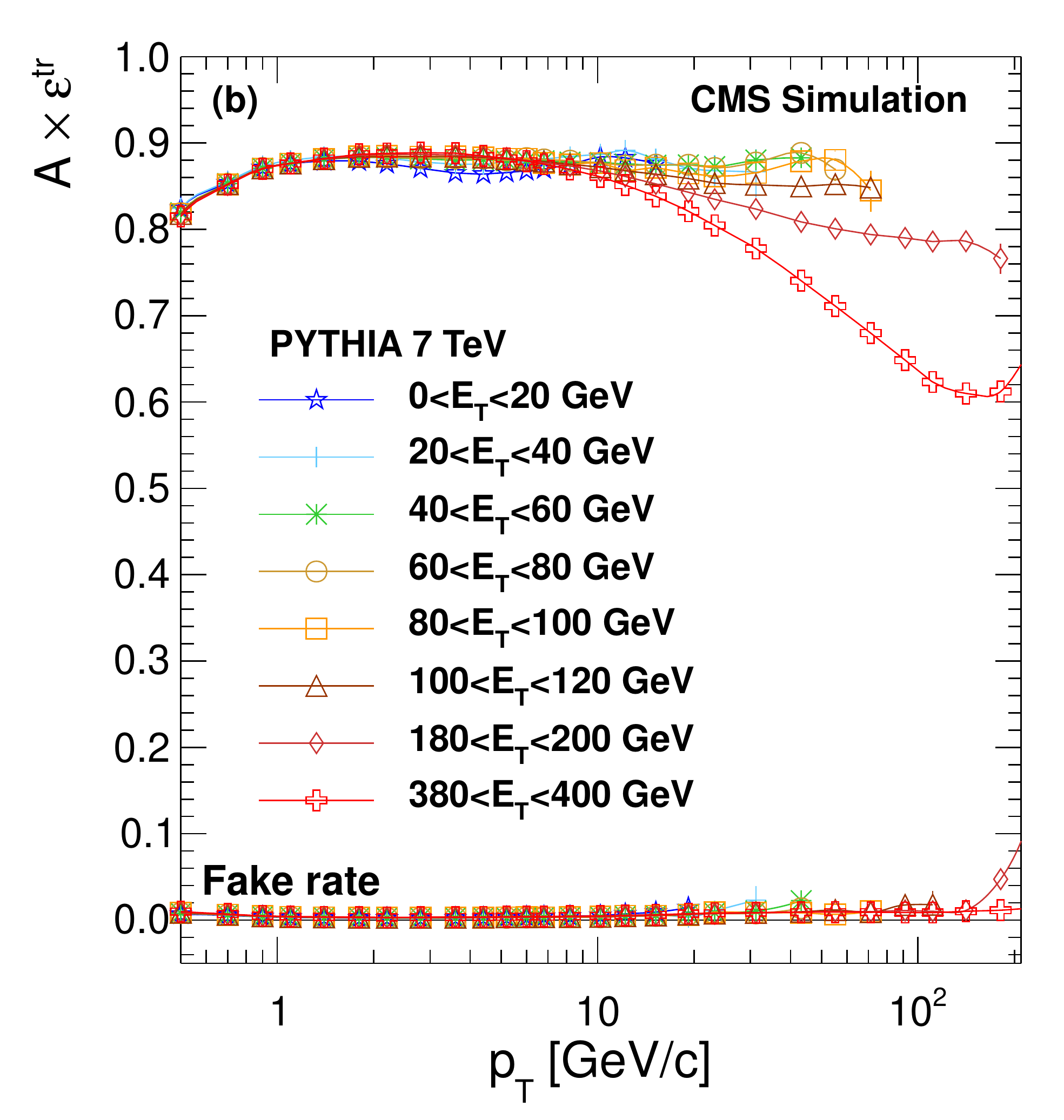}
	\label{fig:occTrk}}
	\caption{(a) The algorithmic tracking efficiency for two different momentum ranges as a function of $\eta$.
	(b) The product of geometrical acceptance (A) with tracking efficiency ($\varepsilon^{\mathrm{tr}}$) (upper points) and
	the misidentification (`fake') rate (lower points)
	as a function of transverse momentum for tracks with $|\eta|<1$ in bins of corrected leading-jet transverse energy.}
	\vspace{4mm}
\end{figure}

With the above-mentioned selections applied to the reconstructed tracks, the algorithmic efficiency determined from simulated
\textsc{pythia} events is greater than 85\% (80\%) for tracks with transverse momentum above 2.0 (0.4)\GeVc\ averaged
over $|\eta|<2.4$ (Fig.~\ref{fig:trkEff}).
In the same kinematic region, misidentified and non-primary tracks are each below 1\%, while multiple reconstruction occurs for less than
0.01\% of tracks.

\section{Event Classification by Leading-Jet Energy}
\label{sec:jetet}

All events in this analysis are classified according to the transverse energy of the most energetic reconstructed jet,
defined as the leading jet.
Jets are reconstructed from calorimeter deposits alone using the anti-\kt\ algorithm~\cite{Cacciari:2008gp} with
cone radius $R=\sqrt{(\Delta\phi)^2+(\Delta\eta)^2}=0.5$.
The measured energy of the jet is adjusted according to corrections based on a MC description of the CMS calorimeter
response with a 3--6\% uncertainty on the jet energy scale~\cite{JME-10-010}.

The motivation for classifying events according to the leading-jet transverse energy is twofold.
First, the degrading effect of the local-track density on the high-\pt\ tracking performance (e.g., inside a jet)
can be parametrised according to this variable.
Based on events simulated with \textsc{pythia} in minimum bias and QCD samples with various thresholds on the hard-scattering
scale ($\hat{p}_{\mathrm{T}}$), the efficiency and misidentification rates of the selected tracks are estimated as a function of
transverse momentum in bins of leading-jet transverse energy (see Fig.~\ref{fig:occTrk}).
Second, as discussed in Section~\ref{sec:evtSel}, calorimeter-based triggers with leading-jet transverse energy thresholds of 15\GeV (Jet15U)
and 50\GeV (Jet50U) were used to extend the \pt\ reach of the 7\TeV measurement.

\begin{figure}[t]
	\centering
	\subfigure{
		\includegraphics[width=0.45\textwidth]{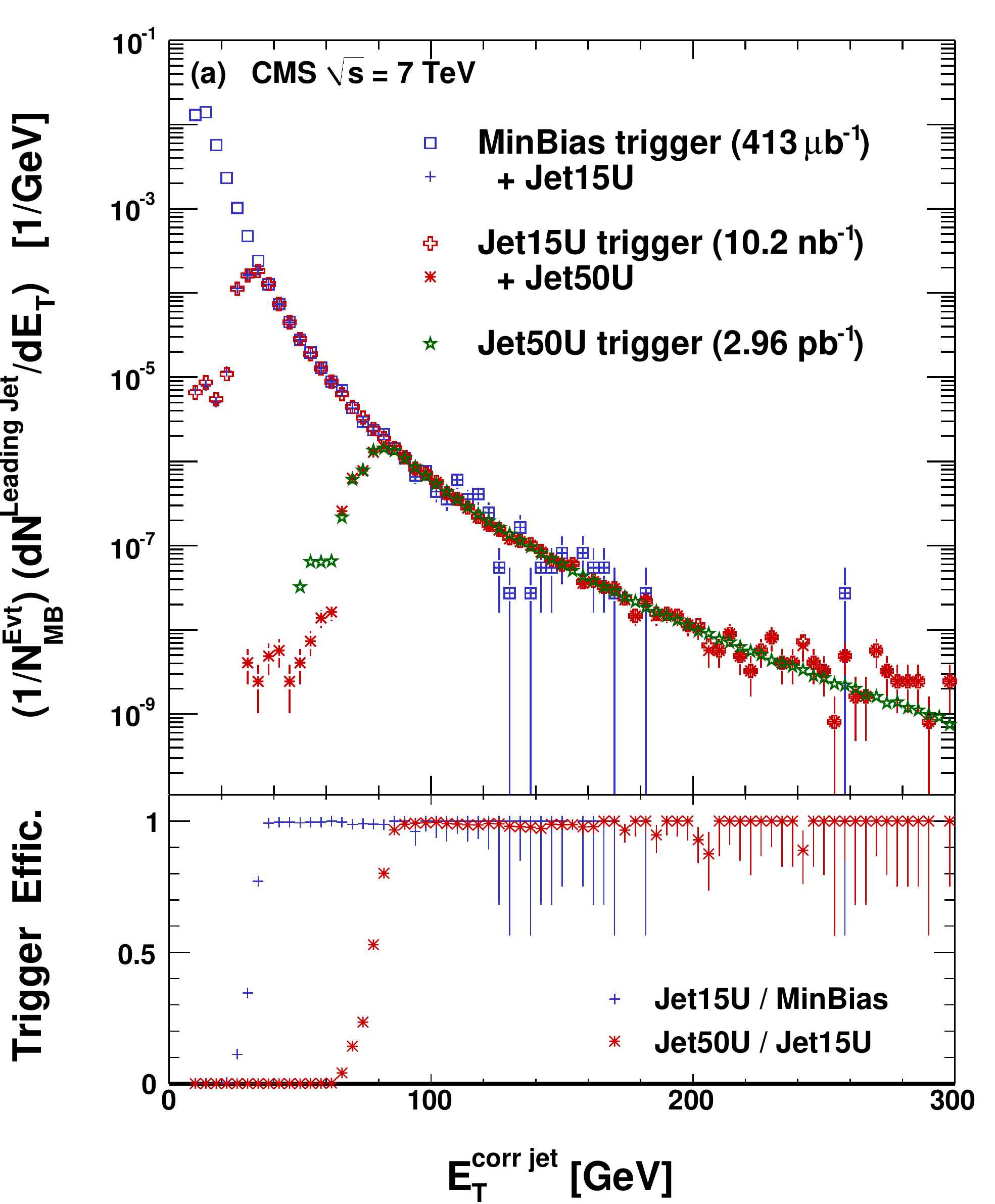}
		\label{fig:hltJetTurnon}}
	\subfigure{
		\includegraphics[width=0.45\textwidth, viewport=0 25 567 719]{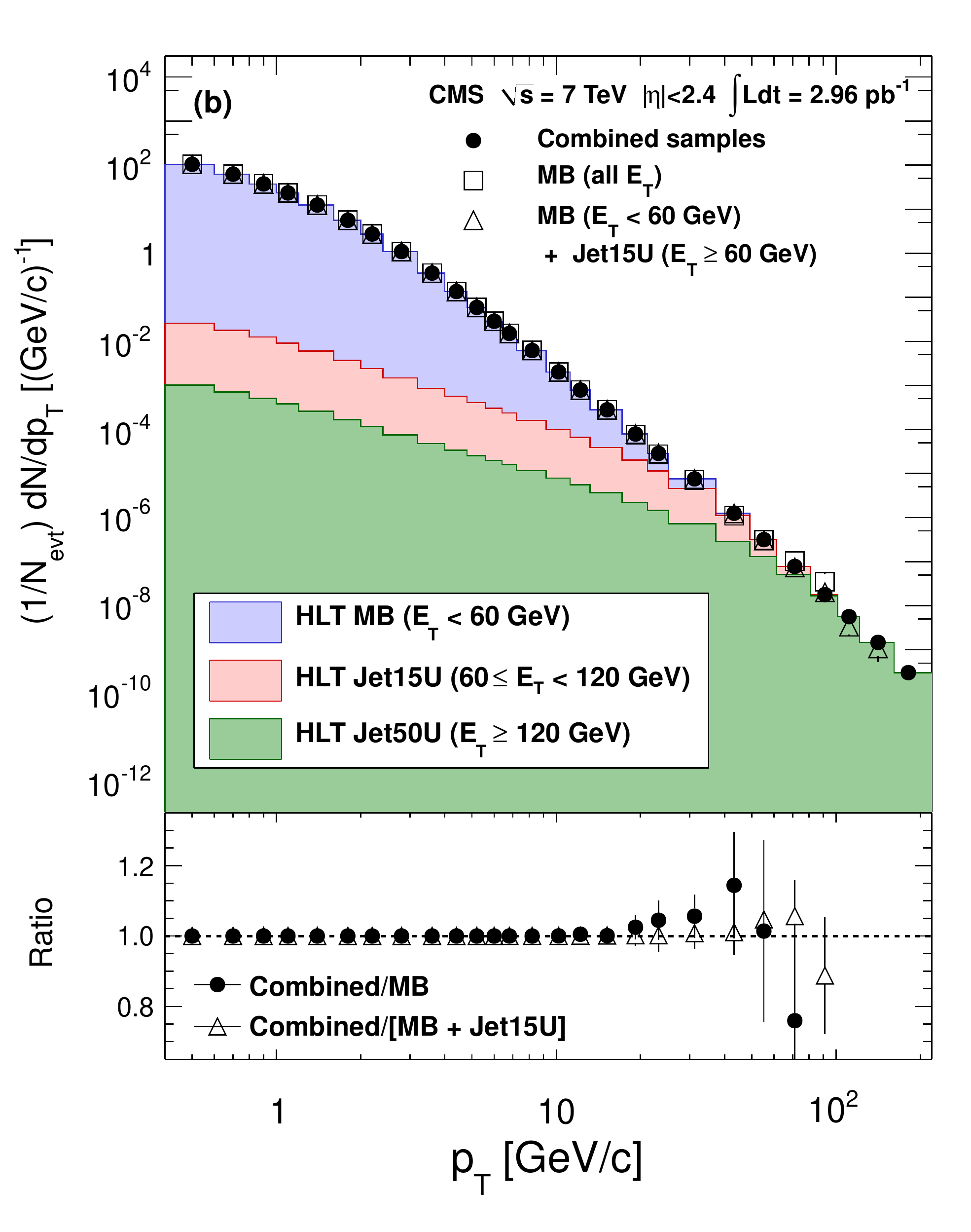}
		\label{fig:spectraLeadingJetBins}}		
		\caption{(a) Upper panel: distributions of the corrected transverse energy of leading jets normalised by the number of selected minimum
		bias events $N^{\mathrm{Evt}}_{\mathrm{MB}}$.
		Lower panel: the efficiency turn-on curves for the jet triggers with uncorrected energy thresholds of 15 and 50\GeV.
        (b) Upper panel: the three contributions to the charged particle transverse momentum spectrum and their sum (solid circles).
        Open squares show the minimum bias spectrum for all values of leading-jet \et; open triangles show the spectrum with
        the addition of only the lower threshold jet trigger.
        Lower panel: the ratio of the combined spectrum to minimum bias only (solid circles) and with the addition of
        only the lower threshold jet trigger (open triangles).}
        	\vspace{4mm}
\end{figure}

To avoid potential biases from the jet-trigger selection, it is desirable to operate in a region where the trigger is fully efficient.
The region above which the jet trigger with an uncorrected energy threshold of 15\GeV becomes fully efficient
is determined by first plotting the leading-jet \et\ distribution for a sample of events selected with the prescaled minimum bias trigger
and the offline selections described in Section~\ref{sec:evtSel}.  This distribution is then compared to the subset of those events which
also fire the 15\GeV jet trigger as a function of corrected transverse energy.  The resulting ratio is the trigger efficiency curve presented in
the lower panel of Fig.~\ref{fig:hltJetTurnon}.  The 15\GeV jet trigger achieves more than 99\% efficiency at a corrected energy of $\et=45$\GeV.
The analogous procedure is repeated on a sample of events selected by the 15\GeV jet trigger
to determine that the 50\GeV jet trigger becomes fully efficient above $\et= 95$\GeV.
For the trigger efficiency study, an early subset of the data (10.2\nbinv) was used, because the minimum bias and
lower-threshold jet triggers were highly prescaled in the later runs.
In the upper panel of Fig.~\ref{fig:hltJetTurnon}, the $\et$ distributions from the jet-triggered sample are normalised per
equivalent minimum bias event by matching their integrals in the regions where the triggers are fully efficient.

For the 7\TeV analysis, events are divided into three classes based on leading-jet \et:
below 60\GeV, between 60 and 120\GeV, and above 120\GeV.
Since each event is uniquely assigned to one such leading-jet \et\ range, the overall $dN_{\mathrm{ch}}/d\pt$ distribution
is simply the sum of the spectra from the three ranges, each corresponding to a fully-efficient HLT selection (i.e., minimum bias,
15\GeV jet trigger, and 50\GeV jet trigger).
The contributions to the spectra from the jet-triggered events are normalised per selected minimum bias event; the fraction of
minimum bias events containing a leading jet with greater than either 60 or 120\GeV is calculated as shown in Fig.~\ref{fig:hltJetTurnon}
by matching the fully-efficient regions of the leading-jet \et\ distributions.
The three contributions to the combined charged particle transverse momentum spectrum are shown in Fig.~\ref{fig:spectraLeadingJetBins}.
The lower panel of that figure compares the combined spectrum first to the minimum bias spectrum alone and then to a spectrum
constructed with the addition of only the lower-threshold jet trigger.  These are all in good agreement within their respective statistical uncertainties.
A \pt-dependent systematic uncertainty of 0--4\% is attributed to the normalisation of the contributions from the triggered samples.
This value is determined by changing the leading-jet \et\ ranges that separate the three samples
(e.g., to $\et = 40$ and 100\GeV), by basing the normalisation directly on the HLT prescale values, and by comparing the normalisations
determined from different subsets of the full data sample.

\section{Corrections and Systematic Uncertainties}
\label{sec:corr}

To obtain the final phase-space-invariant charged particle differential momentum distribution, a number of corrections
must be applied to the raw distributions of reconstructed charged particles, according to the following equation:

\begin{equation}
E \frac{d^{3}N_{\mathrm{ch}}}{dp^{3}} (\pt,\eta) = \frac{\sum_{_{M,\et^{\mathrm{jet}}} } N_{\mathrm{track}}^{\mathrm{raw}}(M,\et^{\mathrm{jet}},\pt,\eta)  \cdot  w_{\mathrm{tr}}(\pt,\eta,\et^{\mathrm{jet}}) \cdot  w_{\mathrm{ev}}(M) }{2\pi \pt \cdot  \Delta \pt \cdot  \Delta\eta \cdot  \sum_{_{M}} N^{\mathrm{selected}}(M) \cdot (1-f^{0}_{\mathrm{NSD}})^{^{-1}} \cdot (1+f^{\mathrm{pileup}}) \cdot  w_{\mathrm{ev}}(M)},
\label{eqn:corrspectra}
\end{equation}

where $N_{\mathrm{track}}^{\mathrm{raw}}$ is the raw number of tracks in a bin with transverse momentum width $\Delta \pt$
and pseudorapidity width $\Delta\eta$, and $N^{\mathrm{selected}}$ is the number of selected events.
An event weight $w_{\mathrm{ev}}$ (see Eq.~(\ref{eqn:evtWeight})) is applied as a function of the multiplicity of
reconstructed charged particles ($M$), while a track weight $w_{\mathrm{tr}}$ (see Eq.~(\ref{eqn:trkWeight})) is applied for each $M$
and leading-jet transverse energy ($\et^{\mathrm{jet}}$), as a function of \pt; the final results are summed over $M$ and $\et^{\mathrm{jet}}$.
The number of selected events is corrected for the fraction of NSD events ($f^{0}_{\mathrm{NSD}}$) that have zero reconstructed
tracks in the tracker acceptance of $|\eta|<2.4$ (about 5\%) and for the pileup event fraction ($f^{\mathrm{pileup}}$).

The multiplicity-dependent event weight $w_{\mathrm{ev}}$ accounts for the efficiency of the event selection
for accepting NSD events ($\varepsilon^{\mathrm{selected}}_{\mathrm{NSD}}$) and for the fraction of SD events ($f^{\mathrm{selected}}_{\mathrm{SD}}$)
that contaminate the selected sample (about 5\% overall):

\begin{equation}
w_{\mathrm{ev}}(M) = \frac{1}{\varepsilon_{\mathrm{NSD}}^{\mathrm{selected}}} (1-f^{\mathrm{selected}}_{\mathrm{SD}}).
\label{eqn:evtWeight}
\end{equation}

The correction factor $w_{\mathrm{tr}}$, by which each track is weighted, is calculated for each bin in transverse momentum, pseudorapidity,
and leading-jet transverse energy.
This factor accounts for the geometric detector acceptance ($A$) and algorithmic tracking efficiency ($\varepsilon^{\mathrm{tr}}$),
as well as the fraction of tracks corresponding to the same, multiply reconstructed charged particle ($D$),
the fraction of tracks corresponding to a non-primary charged particle ($S$),
and the fraction of misidentified (`fake') tracks that do not correspond to any charged particle ($F$):

\begin{equation}
w_{\mathrm{tr}}(\pt,\eta,\et^{\mathrm{jet}}) = \frac{(1-F) \cdot (1-S)}{A \cdot \varepsilon^{\mathrm{tr}} \cdot (1+D) }.
\label{eqn:trkWeight}
\end{equation}

The common uncertainty related to the triggering and event selection efficiency is discussed in detail in
Ref.~\cite{Khachatryan:2010us}.  Contributions from uncertain diffractive-event fractions and detector
inefficiencies in the BSC and HF combine to contribute a scale error of $\pm 3.5$\% to the total systematic
uncertainty at $\sqrt{s}=7$\TeV (see Table~\ref{tab:sys}).  At $\sqrt{s}=0.9$\TeV, the diffractive fractions are slightly better constrained,
hence an uncertainty of $\pm 3.2$\% is assigned.

Using simulated events generated with \textsc{pythia} tune D6T, the various terms in Eq.~(\ref{eqn:trkWeight}) are estimated by matching selected
reconstructed tracks to simulated tracks based on the requirement that they share 75\% of their hits.
As an example, the algorithmic efficiency ($\varepsilon^{\mathrm{tr}}$) versus $\eta$ is presented in Fig.~\ref{fig:trkEff}.  The slight asymmetry between
the positive and negative hemispheres is attributed to a slightly displaced beam-spot and the distribution of dead channels in the tracker.
The systematic uncertainties assigned to the various tracking corrections are discussed below and are summarised,
along with the total systematic uncertainty, in Table~\ref{tab:sys}.

\begin{table}[tb]
\caption{Summary of the various contributions to the estimated systematic uncertainty. }
\centering
\begin{tabular}{ l c c}
\\ \hline \hline
Source & \multicolumn{2}{c}{Uncertainty [\%]} \\
Collision energy & 0.9\TeV & 7\TeV \\
\hline
Event selection  & 3.2 & 3.5 \\                    % same as quoted in QCD-10-006
Pileup effect on vertexing & 0.2 & 1.2 \\                      % largest correction for pileup with abs(dz)<1cm
Acceptance  & 1.5 & 1.5 \\                          % pixel clusters (0.3%) + varying generator tune (1%) + beam-spot x,y,z position (1%)
Reconstruction efficiency & 2.2 & 2.2 \\                       % varying track selection cuts (1%) + varying generator tune (2%)
Occupancy effect on efficiency & 0.0--0.5 & 0.0--2.8 \\       % convolution of 10% jet energy uncertainty with factorized dependence
Misidentified track rate & 0.3--1.0 & 0.3--3.0 \\                              % 30% of fake rate based on different generators
Correction for secondary particles & 1.0 & 1.0 \\          % 100% of secondary rate (known shortcomings of K0 and lambda in generators)
Momentum resolution and binning & 0.3--1.5 & 0.3--2.7 \\       % varying form of fit to dN/dpt (various power laws and exponentials)
Normalisation of jet-triggered spectra & -- & 0.0--4.0 \\  % pr9 vs. m6rr (3.3%). consistent with  ET cut  60->40 GeV,  avg prescale from lumiCalc
\hline
Total & 4.3--4.7 & 4.7--7.9 \\
Total excluding event selection uncertainty & 2.9--3.4 & 3.1--7.1 \\
Total including luminosity uncertainty & 11.4--11.6 & 5.1--8.1 \\  %luminosity error of 4% for 7 TeV and 11% for 0.9 TeV
\hline \hline
\vspace{2mm}
\end{tabular}
\label{tab:sys}
\end{table}

The uncertainty on the geometrical acceptance of the tracker was estimated from three sources.  First, the efficiency
of the pixel hit reconstruction was estimated from a data-driven technique involving the projection of two-hit combinations
(called tracklets) onto the third layer in search of a compatible hit.  The observed efficiency of $(99.0\pm0.5)$\% leads to
a 0.3\% uncertainty on the acceptance of pixel-seeded tracks.  Second, the variation of the geometrical acceptance
was estimated for a variety of generator tunes including \textsc{pythia8} \cite{Sjostrand:2007gs} and the \mbox{Perugia0} \cite{Skands:2009zm} tune of \textsc{pythia}.
Third, the variation was estimated after shifting the generated beam-spot and modifying the width of the generated $z$ vertex
distribution.  The latter two effects each contribute a 1\% shift in the acceptance.

In a similar fashion, using the different generator tunes results in a 2\% shift in the reconstruction efficiency.  An additional series of checks was
performed by varying the cuts imposed during the track selection and in the determination of the corresponding MC-based corrections.
The resulting variation in the corrected results contributes another 1\% to the reconstruction efficiency uncertainty.

Since the dependence of the reconstruction efficiency on local hit density has been parametrised in terms of leading-jet transverse energy,
both the uncertainty on the jet energy scale and the accuracy of the jet-fragmentation description become relevant.
The former contribution is estimated by convolving the dependence of the tracking efficiency on the leading-jet transverse energy
(see Fig.~\ref{fig:occTrk}) with a 4\% uncertainty in the jet energy scale~\cite{JME-10-010}.  The latter contribution is estimated by comparing
the \textsc{pythia}-based corrections to \textsc{herwig++}~\cite{Bahr:2008pv}.
The resulting \pt-dependent uncertainty on the occupancy is in the range (0.0--2.8)\%.

Based on studies of different generator tunes and MC samples with different hard-scattering scales, the assigned uncertainty
to the misidentified-track correction grows linearly as a function of \pt\ from 0.3 to 3.0\%.
An additional check was performed for tracks with \pt\ above 10\GeVc\ to correlate the reconstructed track momentum with
the deposited energy in the projected ECAL and HCAL cells.
For the selected tracks in this analysis, there is no evidence of any excess of high-\pt\ misidentified tracks characterised by
atypically little energy deposited in the calorimeters.
The correction for secondaries and feed-down from weak decays is assigned a
1\% systematic uncertainty, which is large compared to the scale of the contributions, but intended to account for the uncertainties
in the \PKzS\ and \PgL\ fractions~\cite{Khachatryan:2011tm}.

The tendency for finite bin widths (up to 40\GeVc) and a finite transverse momentum resolution
(rising from 1 to 5\% in the range \pt = 10--150\GeVc)
to deform a steeply falling spectrum is corrected based on the shape of the \pt\ spectrum and the MC-based \pt\ response matrix.
The effect of momentum resolution alone is 0.5--2.5\%, while the wide binning results in an additional
correction ranging from a fraction of a percent up to approximately 20\% in the widest high-\pt\ bins.
The correction for the two effects is determined by fitting an empirical function to the differential yield, smearing it with the MC-based
momentum resolution, re-binning into the bins of the final invariant yield, and dividing by the original fitted form.
The quoted systematic uncertainty of 0.3--2.7\% is estimated by varying the fitted form of the spectrum and by performing multiple
iterations of the unsmearing with successively more accurate input spectra.

In addition to the uncertainties from the event selection efficiency weighting and the tracking corrections described above,
the total systematic uncertainty contains a contribution
from the uncertainty on the estimation of the event pileup fraction of 0.2 and 1.2\% for the 0.9 and 7\TeV data, respectively.
In the cases where the total integrated luminosity is used to normalise the results,
this contributes an additional 4\% (11\%) scale uncertainty~\cite{EWK-10-004,EWK-11-001} for $\sqrt{s}$ = 7 (0.9)\TeV.
Assuming that the various \pt-dependent contributions are uncorrelated, the total systematic uncertainty is determined
from their sum in quadrature, as indicated in Table~\ref{tab:sys}.

\section{Results}
\label{sec:results}

\begin{figure}[t]
	\centering
		\subfigure{
		\includegraphics[width=0.45\textwidth]{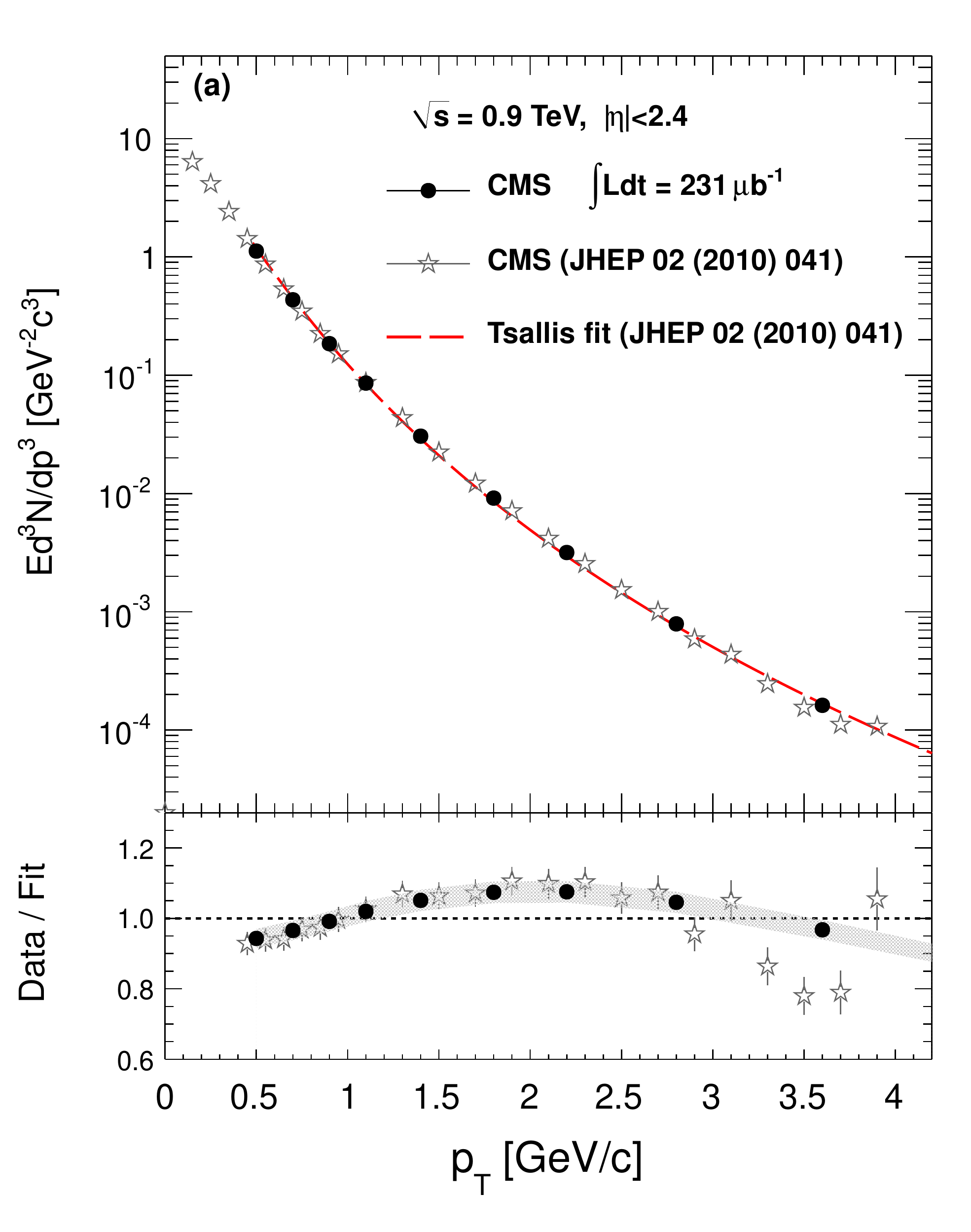}
		\label{fig:spectraCMS900}}
	\subfigure{
		\includegraphics[width=0.45\textwidth]{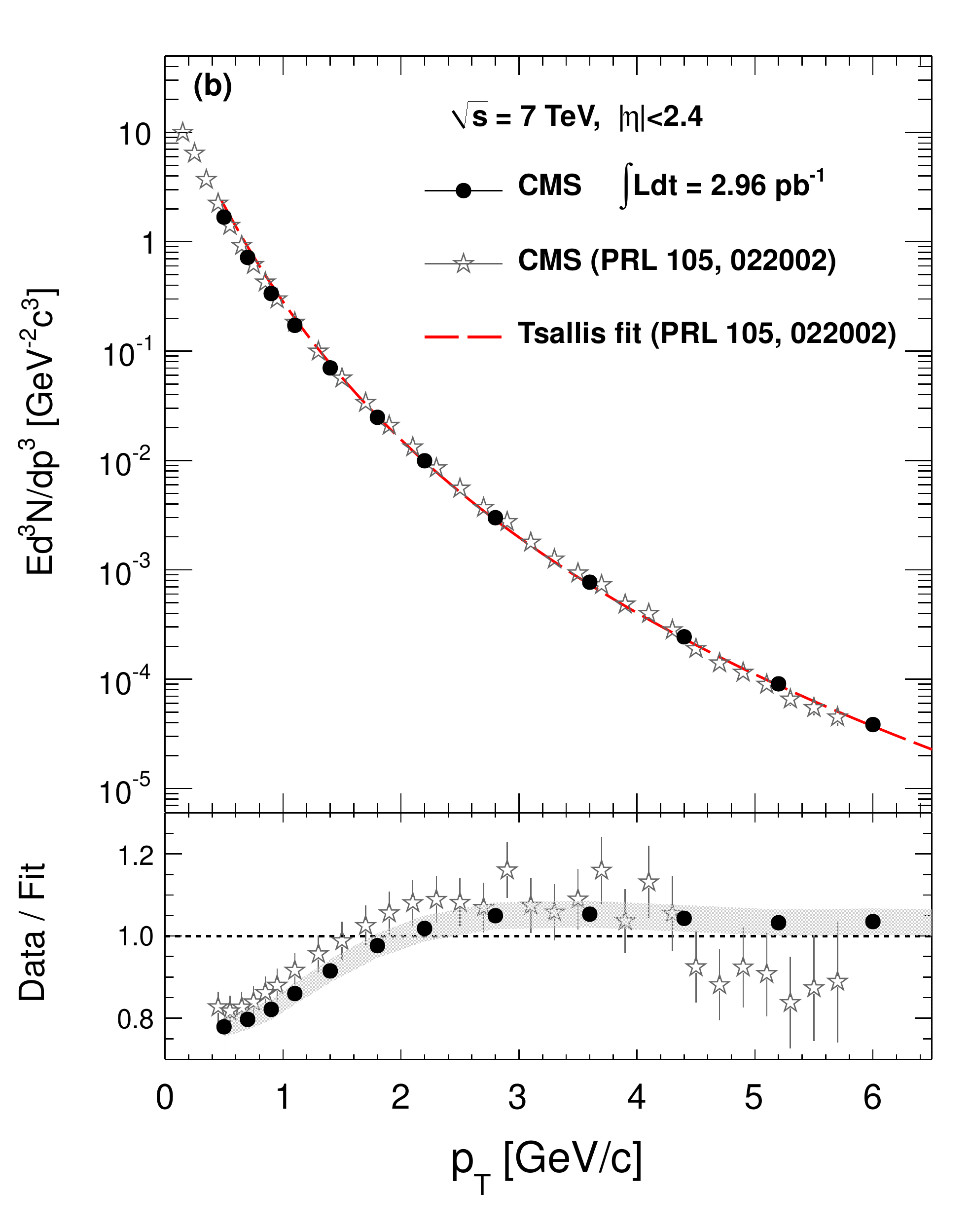}
		\label{fig:spectraCMS}}
	\caption{(a) Upper panel: the invariant charged particle differential yield from the present analysis (solid circles) and the previous CMS
	measurements at $\sqrt{s}=0.9$\TeV (stars) over the limited \pt\ range of the earlier result.
	Lower panel: the ratio of the new (solid circles) and previous (stars) CMS results to a Tsallis fit of the earlier measurement.
	Error bars on the earlier measurement are the statistical plus systematic uncertainties added in quadrature.
	The systematic uncertainty band around the new measurement consists of all contributions,
	except for the common event selection uncertainty.  (b) The same for $\sqrt{s}=7$\TeV.}
	\vspace{4mm}
	\label{fig:spectraCMSboth}
\end{figure}

After applying the corrections described in the previous section, the resulting invariant differential yields for charged particles within $|\eta|<2.4$
are shown for a limited \pt\ range in Figs.~\ref{fig:spectraCMS900} and~\ref{fig:spectraCMS} in order to quantify the agreement with
previous CMS measurements at $\sqrt{s}=$~0.9 and 7\TeV~\cite{Khachatryan:2010xs,Khachatryan:2010us}.
At each energy, both CMS measurements are divided by a Tsallis fit~\cite{tsallis} to the earlier measurement and the ratios compared in the lower
panels.  For the earlier measurements, the error bars indicate the statistical plus systematic uncertainties added in quadrature.
The bands around the new measurements represent all contributions to the systematic uncertainty, except the contribution
from the common event selection.  Statistical uncertainties are negligible on the new measurements in this \pt\ range.
Below $\pt=4$\GeVc\ for the 0.9\TeV sample and below $\pt=6$\GeVc\ at $\sqrt{s}$ = 7\TeV, which are the limits of the
previously published CMS spectra, the new results are in reasonable agreement with the earlier measurements.  However, the measured spectra
do deviate from the Tsallis fits in the earlier papers by as much as 20\% at low \pt.
The origin of the small difference between the two CMS measurements at $\sqrt{s}=7$\TeV
is attributed to the different tracking algorithms used in the two measurements, as well as the different \textsc{pythia}
tunes used to determine the tracking corrections.

\begin{figure}[t]
	\centering
	\subfigure{
		\includegraphics[width=0.45\textwidth]{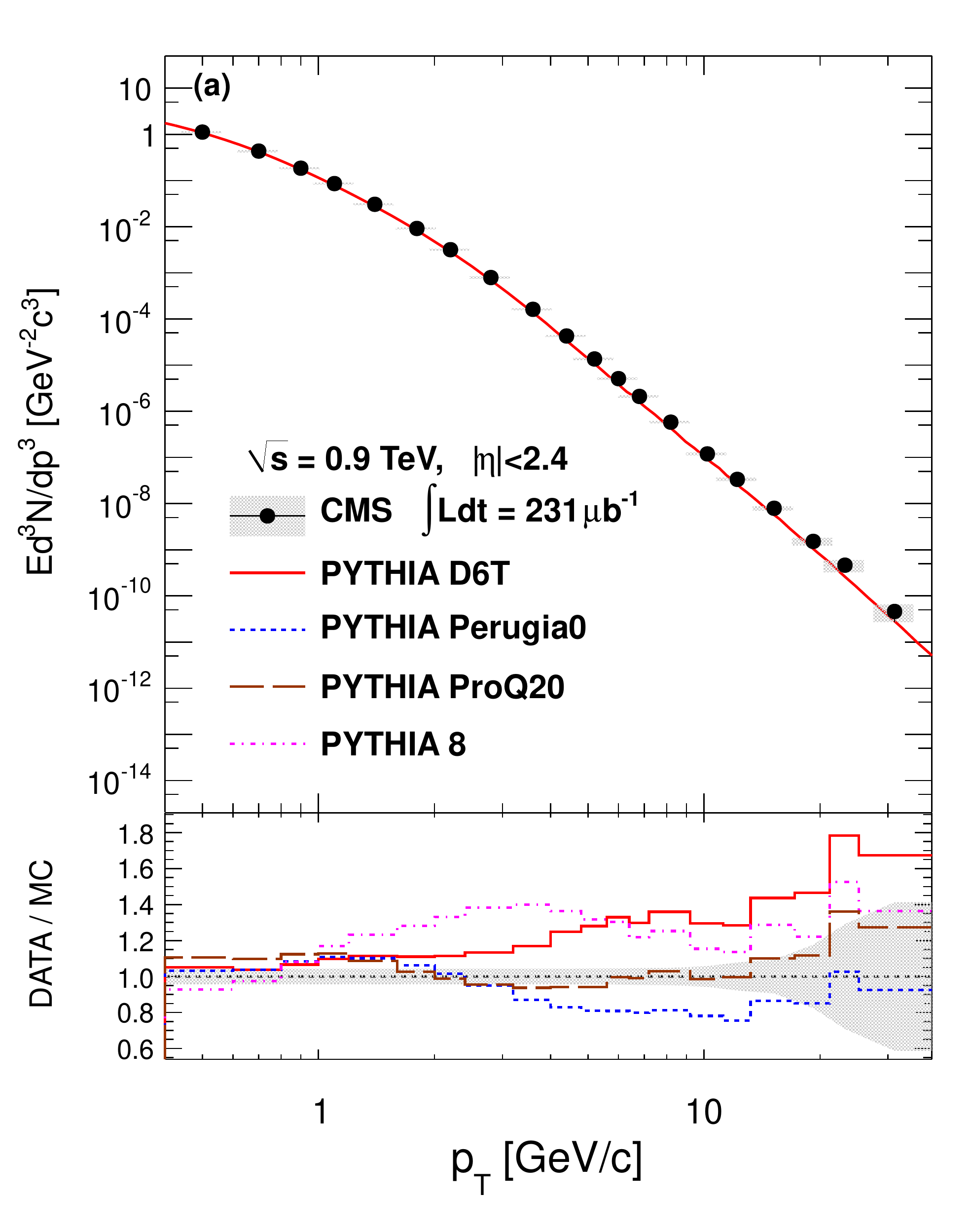}
		\label{fig:spectraGEN900}}
	\subfigure{
		\includegraphics[width=0.45\textwidth]{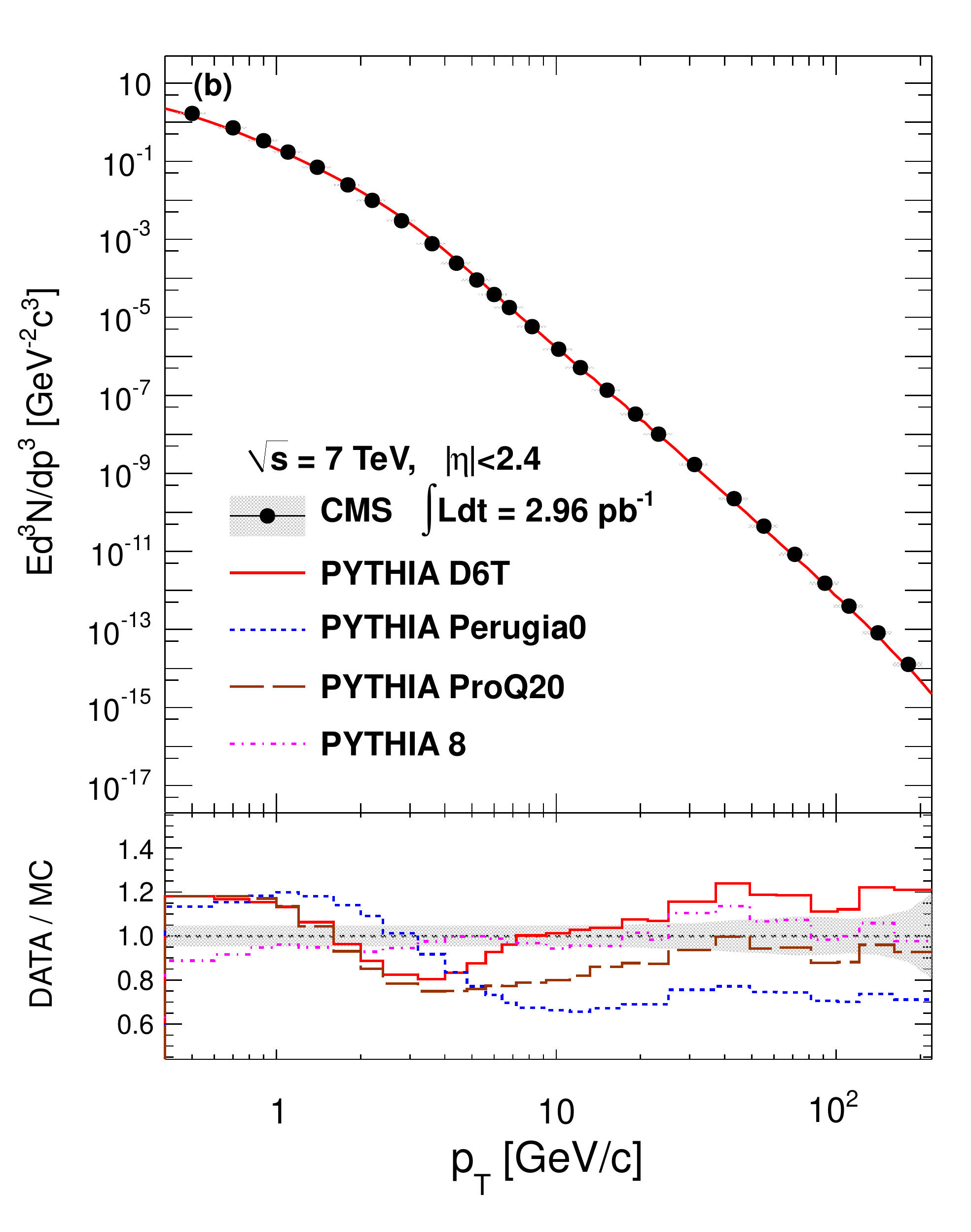}
		\label{fig:spectraGEN}}
	\caption{(a) Upper panel: the invariant charged particle differential yield at $\sqrt{s}=0.9$\TeV compared with the predictions of
	four tunes of the \textsc{pythia} MC generator.
	Lower panel: the ratio of the new CMS measurement to the four \textsc{pythia} tunes.  The grey band corresponds
	to the statistical and systematic uncertainties added in quadrature.  (b) The same for $\sqrt{s}=7$\TeV.}
	\vspace{4mm}
        \label{fig:spectra900GeVn7TeV}
\end{figure}

In the upper plots of Figs.~\ref{fig:spectraGEN900} and~\ref{fig:spectraGEN}, the charged particle differential transverse
momentum yields from this analysis are displayed for $\sqrt{s}$ = 0.9 and 7\TeV, respectively. The latter distribution covers
the \pt\ range up to 200\GeVc, the largest range ever measured in a colliding beam experiment. Also shown in the figures
are various generator-level MC predictions for the yields \cite{Bartalini:2009xx,Sjostrand:2007gs,Skands:2009zm,Buckley:2009bj}.
The lower plots of Figs.~\ref{fig:spectraGEN900} and~\ref{fig:spectraGEN} show the ratios of the data to the various MC predictions.
As already observed in Ref.~\cite{Khachatryan:2010us}, there is a deficit of $\pt < 1$\GeVc\ particles in the
predicted 7\TeV spectra for several of the popular \textsc{pythia} tunes.
For the whole \pt\ range above 1\GeVc, \textsc{pythia8} is the most consistent with the new 7\TeV result (within 10\%).
This provides an important constraint on the different generator parameters responsible for sizable variations among the tunes.
A similar but slightly larger spread is observed in Fig.~\ref{fig:spectraGEN900} for different generator parameters at $\sqrt{s} =0.9$\TeV,
where the CMS measurement is most consistently described by the ProQ20 tune.

\begin{figure}[t]
        \centering
                \subfigure{
                \includegraphics[width=0.45\textwidth]{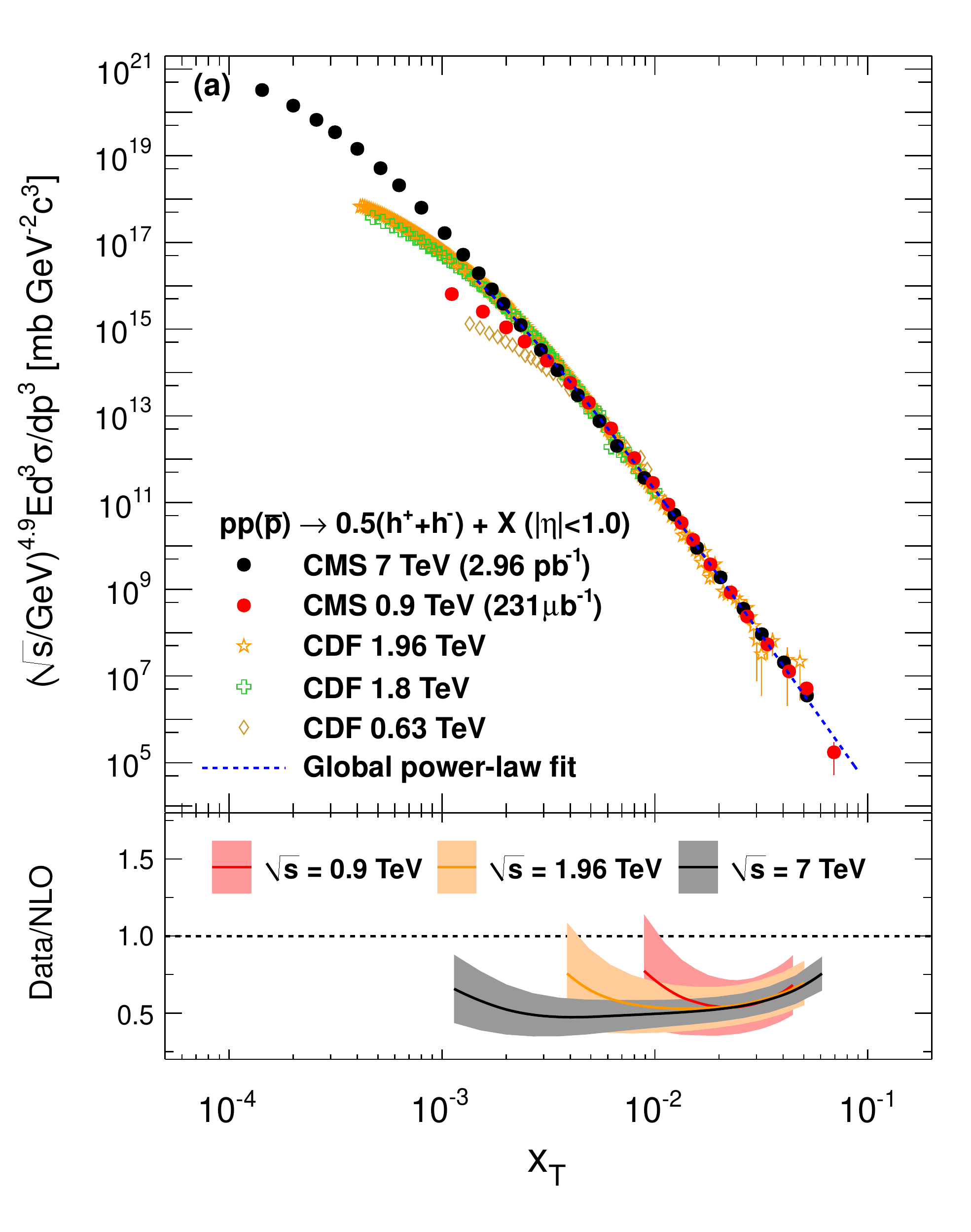}
                \label{fig:xtScaling}}
        \subfigure{
                \includegraphics[width=0.45\textwidth]{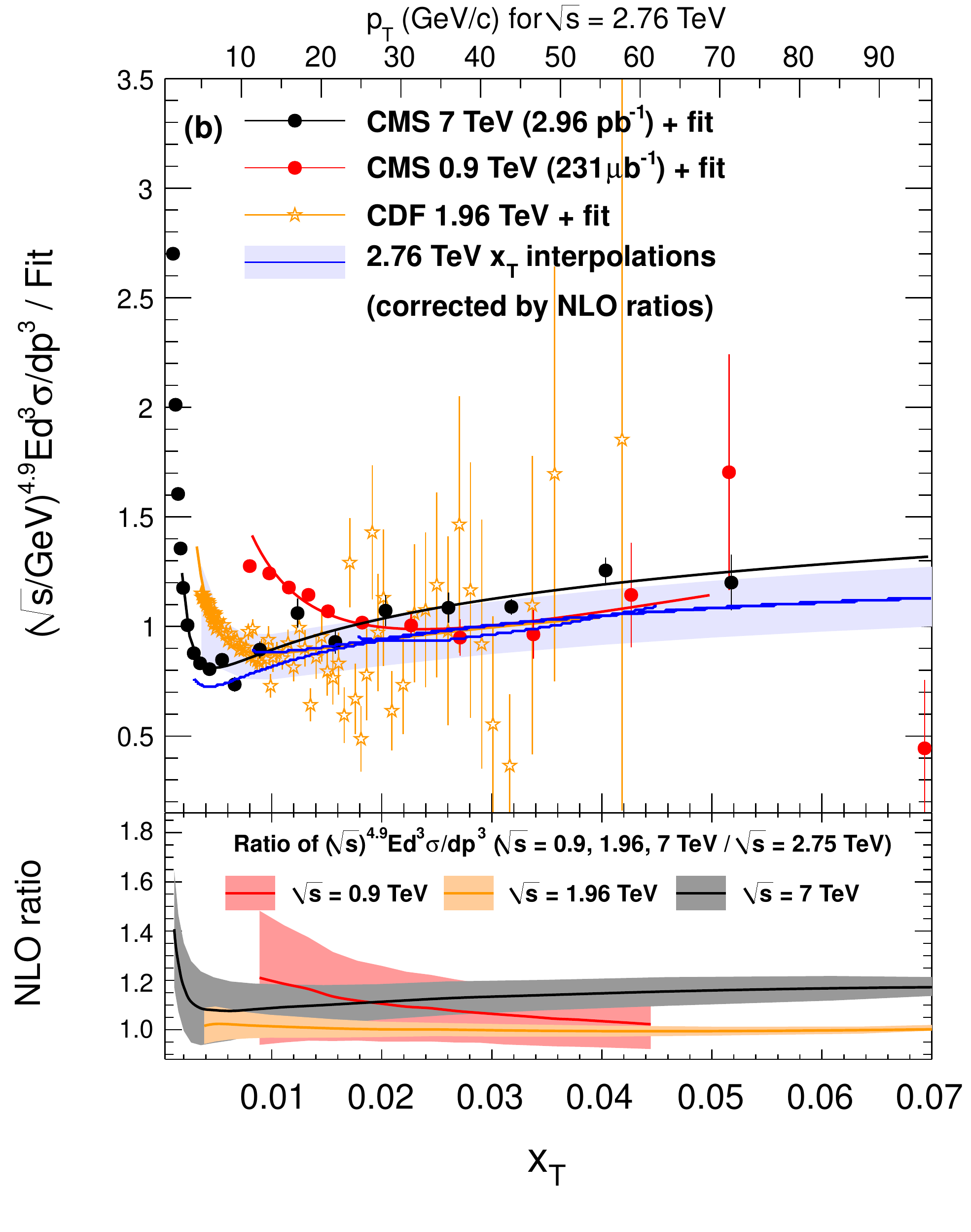}
                \label{fig:xtScalingRatio}}
        \caption{
	(a) Upper panel: inclusive charged particle invariant differential cross sections, scaled by $\sqrt{s}^{4.9}$, for $|\eta|<1.0$
        as a function of the scaling parameter \xt. The result is the average of the positive and negative charged particles.
	Lower panel: ratios of differential cross sections measured at 0.9, 1.96, and 7 TeV to those predicted by NLO calculations for 
        factorisation scales ranging from 0.5--2.0 \pt.  
	(b) Upper panel: ratios of the scaled differential cross sections to the global power-law \xt\ fit described in the text (coloured markers)
	and fits to these ratios (similarly coloured thin lines).
	The expected ratio for $\sqrt{s}=2.76$\TeV after applying NLO-based corrections to each of the three measurements
	as described in the text (solid blue lines).  The uncertainty from the NLO parameters is represented by the shaded band.
	The upper axis translates \xt\ to \pt\ for $\sqrt{s}=2.76$\TeV.
	Lower panel: ratios of the NLO-calculated cross sections at three different energies, scaled by $\sqrt{s}^{4.9}$, to the cross section
	calculated at $\sqrt{s}=2.75$\TeV.  The width of the bands represents the variation of the factorisation scale by a factor of two.}
	\vspace{4mm}
\end{figure}

As discussed in Ref.~\cite{Arleo:2009ch,Arleo:2010kw}, a robust prediction of pQCD hard processes is the power-law scaling of the inclusive
charged particle invariant differential cross section with the variable \xt:

\begin{equation}
E\frac{d^3\sigma}{dp^3} = F(\xt)/\pt^{n(\xt,\sqrt{s})} =   F'(\xt)/\sqrt{s}^{n(\xt,\sqrt{s})},
\label{eqn:xtscaling}
\end{equation}

where $F$ and $F'$ are independent of $\sqrt{s}$, and the slow evolution of the power-law exponent $n$ with \xt\ and
$\sqrt{s}$ ($n\simeq$ 5--6) is due to the running of $\alpha_{\mathrm{s}}$ and changes in the  parton distribution and fragmentation functions.
In the upper plot of Fig.~\ref{fig:xtScaling}, the 0.9 and 7\TeV pp measurements from this
analysis are compared to the empirical scaling observed from measurements
over a range of lower p$\bar{\mathrm{p}}$ collision energies by plotting \mbox{$\sqrt{s}^{n} \, E \,d^{3}\sigma / dp^{3} $}.
For the purpose of reporting the CMS results as differential cross sections, the integrated luminosities
for the analysed data samples were measured according to the descriptions in Ref.~\cite{EWK-10-004,EWK-11-001}.
Also, to compare with the published results from the CDF experiment at $\sqrt{s}=0.63$, 1.8, and 1.96\TeV, the pseudorapidity
range has been restricted to $|\eta|<1.0$.
Whereas an exponent $n=5.5$ was found in Ref.~\cite{Arleo:2010kw} from a global fit to only the previous p$\bar{\mathrm{p}}$ measurements from
$\sqrt{s}=0.2$ to 1.96\TeV, the \xt\ scaling presented in this paper is optimised for use in an interpolation between the CDF and CMS
measurements from $\sqrt{s}=0.9$ to 7\TeV.  Within this range, the best scaling is achieved with an exponent
of $n=4.9\pm0.1$.  This is consistent with the predictions of next-to-leading-order (NLO) calculations,
where the scaling is also found to be optimised for this value of the exponent~\cite{Arleo:2010kw}.
From the lower panel of Fig.~\ref{fig:xtScaling}, it is apparent that the NLO calculations over-predict the measured cross sections by almost 
a factor of two at all collision energies.  This is in spite of the relatively good agreement in the inclusive jet spectrum~\cite{Chatrchyan:2011me,Aad:2010wv}, 
which suggests that the fragmentation functions are not well tuned for LHC energies.

The CMS results are consistent over the accessible \xt\ range with
the empirical \xt\ scaling given by Eq.~(\ref{eqn:xtscaling}) and established at lower energies.
This quality of the scaling is more easily seen in the upper panel of Fig.~\ref{fig:xtScalingRatio}, where the points show the ratio of the
various differential cross sections, scaled by $\sqrt{s}^{4.9}$, to the result of a global power-law fit to the CDF and CMS
data from Fig.~\ref{fig:xtScaling}.
The fitting function is of the form $F'(\xt) = p_0 \cdot [1+(\xt/p_1)]^{p_2}$, where $p_0$, $p_1$, and $p_2$ are free parameters, and
the region below $\pt=3.5$\GeVc\ has been excluded to avoid complications from soft-particle production.
Considering the somewhat na\"{i}ve power-law function and the expected non-scaling effects~\cite{Stratmann:2010bh},
the new measurement is in reasonable agreement with the global power-law fit result (within roughly 50\%) over its full \xt\ range.

\section{Interpolation to 2.76\TeV}
\label{sec:interpolation}

In order to construct a predicted reference charged particle differential cross section at $\sqrt{s}$ = 2.76\TeV for comparison with
the measured PbPb heavy-ion spectrum, two different techniques are used in partially overlapping transverse momentum regimes.
In the high-\pt\ range from 5.0--200\GeVc, where approximate \xt\ scaling is expected to hold, the estimated 2.76\TeV cross section
is derived from a common \xt-scaling curve, based on the CDF and CMS measurements shown in Fig.~\ref{fig:xtScaling}.
In the low-\pt\ range from 1.0--20\GeVc, it is possible to interpolate directly between the several measured cross section values
as a function of $\sqrt{s}$ at each fixed \pt\ value.

As discussed in the previous section, the upper panel of Fig.~\ref{fig:xtScalingRatio} shows the residual difference from perfect \xt\ scaling
with exponent $n=4.9$ for the 0.9 and 7\TeV CMS measurements and for the 1.96\TeV CDF
measurement~\citep{Aaltonen:2009ne,PhysRevD.82.119903} .
The $\sqrt{s}$ and \xt\ dependence of the residuals are not unexpected, since this behaviour is predicted by NLO calculations.
This can be seen in the lower panel of Fig.~\ref{fig:xtScalingRatio}, which shows the predicted deviation from perfect \xt\ scaling for
calculated NLO cross sections at several collision energies with respect to a reference centre-of-mass energy of 2.75\TeV~\cite{Arleo:2010kw}.
The calculations were performed using the CTEQ66 parton distribution functions~\cite{Nadolsky:2008zw},
DSS fragmentation~\cite{deFlorian:2007aj}, and a factorisation scale $\mu=\pt$~\cite{Arleo:2010kw}.  Taking the magnitude of the
\xt-scaling violation from NLO (ranging from 0--20\%), each of the three measurements in data (i.e., 0.9, 1.96, and 7\TeV) can be
corrected separately to arrive at an expectation for the 2.76\TeV cross section.  The three independent interpolations based on NLO-corrected
\xt\ scaling are shown as solid blue lines in the upper panel of Fig.~\ref{fig:xtScalingRatio}.
The combined `best estimate'  (shown as a shaded band) has an associated uncertainty that covers the deviations of up to 12\% observed by varying the
factorisation scale from $\mu=0.5\,\pt$ to $\mu=2.0\,\pt$ for each of the three collision energies.
The error band is expanded below $\pt\approx8$\GeVc\ to include
the full difference between the 1.96 and 7\TeV results, since the evolution of the spectra below this value ---
corresponding to $\xt=0.0023$ (7\TeV), 0.0082 (1.96\TeV), and 0.018 (0.9\TeV)  --- is no longer consistently described
by \xt\ scaling and the NLO-based corrections.  In addition to the 12\% contribution from the uncertainty on the NLO-based correction,
the final uncertainty on the interpolated cross section has an additional component to account for possible correlations in the luminosity
uncertainty between the three measurements.  This term, taken as equal to the smallest individual uncertainty (4\%), is added in quadrature.

\begin{figure}[t]
\centering
                \subfigure{
                \includegraphics[width=0.45\textwidth]{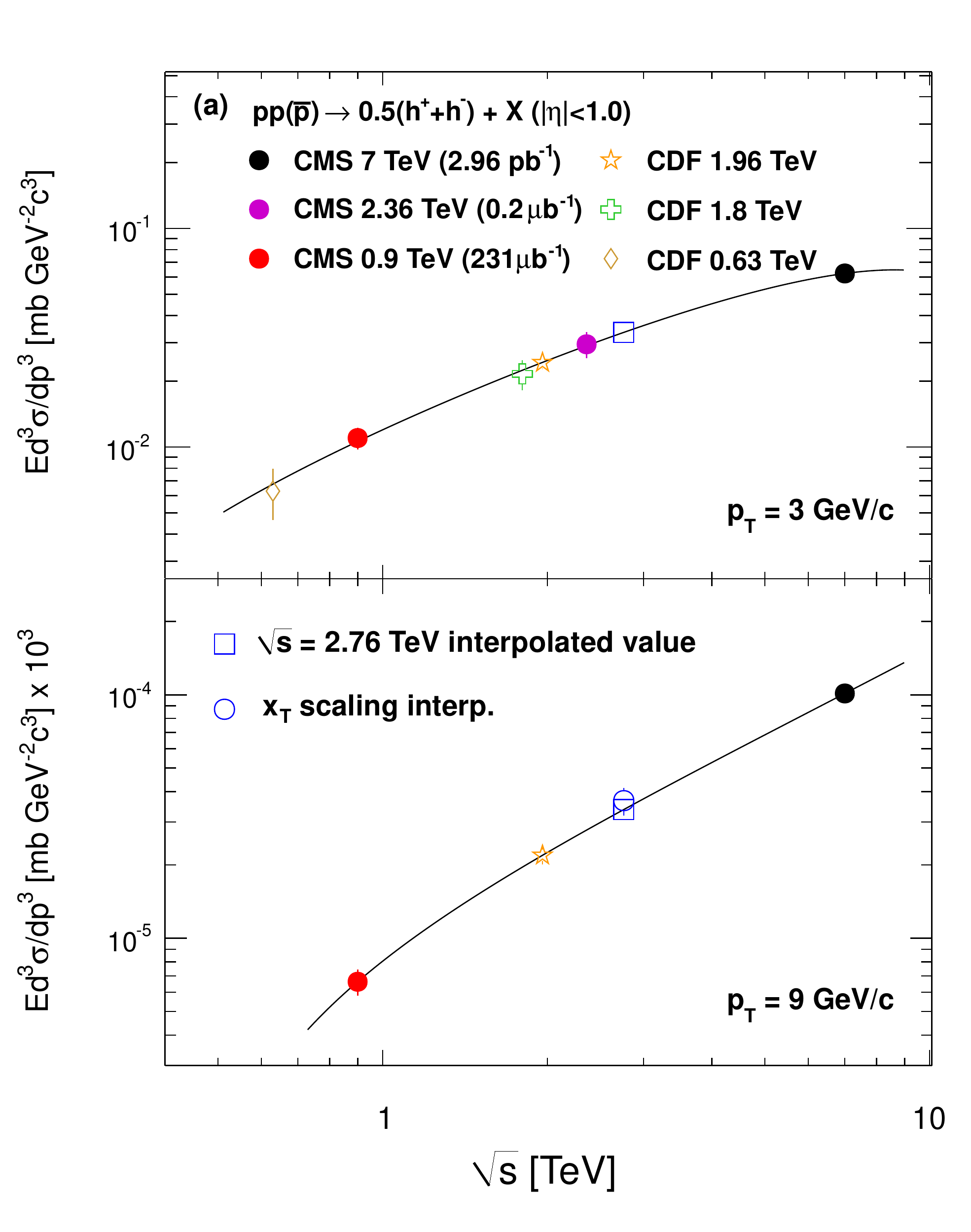}
                \label{fig:ptInterpolation}}
        \subfigure{
                \includegraphics[width=0.45\textwidth]{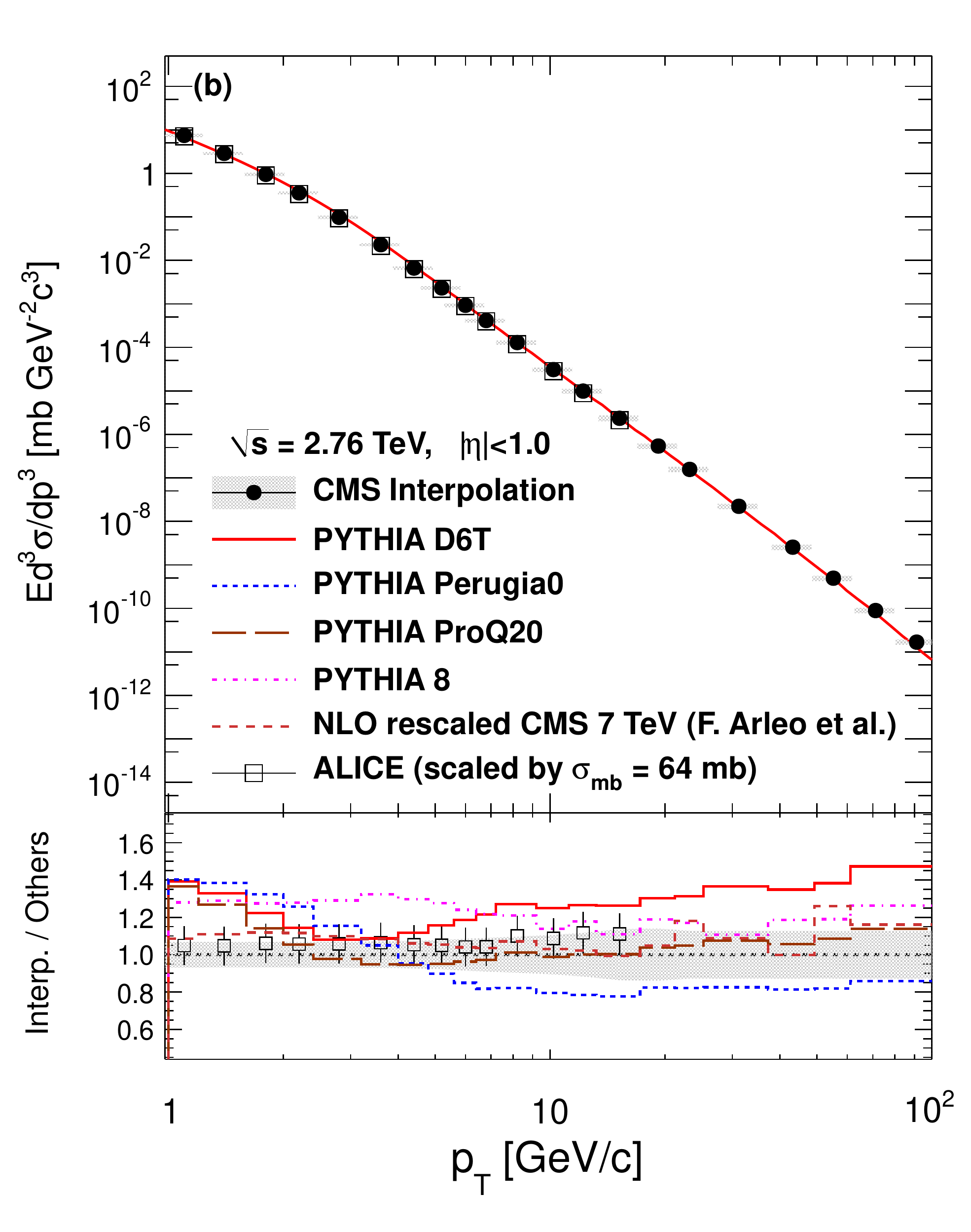}
                \label{fig:interpolation}}
        \caption{
	(a) Interpolations between measured charged particle differential cross sections at different $\sqrt{s}$ for the two example values of $\pt=3$
	and 9\GeVc.  Second-order polynomial fits to the measured data are shown by the solid lines.  The open squares show the resulting interpolated
	cross sections for $\sqrt{s}=2.76$\TeV.
	The open circle on the lower panel represents the corresponding estimate from the \xt-scaling approach in the overlap region where both
	can be estimated.
	(b) Upper panel: the predicted 2.76\TeV charged particle differential transverse momentum cross section,
	based on the combined direct \pt\ interpolation and NLO-corrected \xt-scaling techniques described in the text.
	Lower panel: ratios of combined interpolation to predictions from several \textsc{pythia} tunes,
	an NLO-based rescaling approach~\cite{Arleo:2010kw}, and the \mbox{ALICE} interpolation used in Ref.~\cite{Aamodt:2010jd}.}
	\vspace{4mm}
\end{figure}

The direct interpolation of cross sections at a fixed value of \pt\ is done using CDF measurements at $\sqrt{s}=0.63, 1.8$~and
1.96\TeV~\citep{Abe:1988yu,Aaltonen:2009ne,PhysRevD.82.119903}, the new CMS measurements at $\sqrt{s}=0.9$ and 7\TeV,
as well as an earlier result at $\sqrt{s}=2.36$\TeV~\cite{Khachatryan:2010xs}.  The latter measurement is converted
to a differential cross section assuming the total inelastic cross section of 60.52\,mb from \textsc{pythia}.
At each energy, an empirical fit to the \pt\ distribution is first constructed to provide
a continuous estimation independent of different binning.  Then, in arbitrarily small \pt\ bins, these empirical fits are evaluated and the evolution
of the cross section with $\sqrt{s}$ is parametrised by a second-order polynomial.  Two examples of these fits are shown in
Fig.~\ref{fig:ptInterpolation} for $\pt=3$ and 9\GeVc.  The uncertainty on the value of the fit evaluated at $\sqrt{s}=2.76$\TeV is taken
from the covariance matrix of the fit terms, with an additional 4\% added in quadrature to account conservatively for any correlation in the
luminosity uncertainty between the different measurements.

To arrive at a single interpolated spectrum over the full \pt\ range, a linear combination of the two techniques is used with weights that
vary linearly across the overlap range from $\pt=5$\GeVc\ (only direct interpolation at fixed \pt) to $\pt=20$\GeVc\ (only \xt\
scaling with NLO-based residual correction).  In the \pt\ range where the two techniques overlap, the different methods agree to within their
respective systematic uncertainties.  (The fixed-\pt\ interpolation value is typically around 8\% lower than the \xt\ interpolation.)
The resulting predicted 2.76\TeV differential cross section is shown in the upper panel of Fig.~\ref{fig:interpolation},
and its ratio with respect to various \textsc{pythia} tunes at that centre-of-mass energy in the lower panel.
The uncertainty on the predicted cross section, shown by the grey band in the lower panel, is the weighted sum (where applicable)
of the uncertainties derived from the two methods described in the preceding paragraphs.
Also shown in the lower panel of Fig.~\ref{fig:interpolation} is the ratio of the predicted 2.76\TeV cross section to that found by
simply scaling the CMS measured 7\TeV result by the expected 2.75\TeV to 7\TeV ratio from NLO calculations~\cite{Arleo:2010kw}.
The interpolation used in the recent \mbox{ALICE} publication~\cite{Aamodt:2010jd} is a few percent lower than the result quoted in this paper,
but consistent within the respective systematic uncertainties.
The behavior of the various generators compared to the interpolated 2.76\TeV cross section is broadly similar to the 0.9\TeV invariant yields 
presented in Fig.~\ref{fig:interpolation}. The ProQ20 tune agrees most closely (within 15\%) with the interpolated cross section above 2\GeVc.
Future analysis of a recently recorded 2.76\TeV pp collision sample will provide verification of this result
and a reduction in the systematic uncertainties.

\section{Summary}
\label{sec:summary}

In this paper, measurements of the phase-space-invariant differential yield $E\,d^{3}N_{\mathrm{ch}}/dp^{3}$ at $\sqrt{s}$ = 0.9 and 7\TeV
have been presented for primary charged
particles, averaged over the pseudorapidity acceptance of the CMS tracking system ($|\eta|<2.4$).
The results have been shown to be in reasonable agreement with
the previously published CMS measurements at $\sqrt{s}$ = 0.9 and 7\TeV~\cite{Khachatryan:2010xs,Khachatryan:2010us}
and, except for the surplus of tracks at very low transverse momentum, with \textsc{pythia} leading-order pQCD.
The 7\TeV data are most consistent with \textsc{pythia8}, which agrees at the 10\% level over the full \pt\ range of the measurement.
In contrast, the 0.9\TeV data are considerably better described by the ProQ20 tune.
Additionally, the consistency of the 0.9 and 7\TeV spectra has been demonstrated
with an empirical \xt\ scaling that unifies the differential cross sections from a wide range of collision energies onto a common curve.
Furthermore, within the theoretical uncertainties of the NLO calculations, the residual breaking of \xt\ scaling above $\pt \approx 8$\GeVc\
is consistent between the measured cross sections and the NLO calculations.

This result has removed a large uncertainty from an important ingredient of existing and future PbPb measurements,
namely the pp reference spectrum corresponding to the energy of the 2010 PbPb run: 2.76\TeV per nucleon.
By employing a combination of techniques to interpolate between the results presented here at $\sqrt{s}=0.9$ and 7\TeV, including information from
existing CDF measurements at $\sqrt{s}=0.63$, 1.8, and 1.96\TeV, a pp reference at $\sqrt{s}=2.76$\TeV has been constructed over a large range
of transverse momentum (\pt\ = 1--100\GeVc) with systematic uncertainties of less than 13\%.

\section*{Acknowledgements}
\input{section_acknowledge.tex}

\bibliography{auto_generated}   % will be created by the tdr script.
\cleardoublepage \appendix\section{The CMS Collaboration \label{app:collab}}\begin{sloppypar}\hyphenpenalty=5000\widowpenalty=500\clubpenalty=5000\input{QCD-10-008-authorlist.tex}\end{sloppypar}
\end{document}

%% file: custom-definitions.tex
\newcommand{\xt}{\ensuremath{x_{\mathrm{T}}}\xspace}

\newcommand{\microbinv} {\mbox{\ensuremath{\,\mu\text{b}^\text{$-$1}}}\xspace}

%% file: section_acknowledge.tex
\label{sect:acknowledge}

We wish to congratulate our colleagues in the CERN accelerator departments for the excellent performance of the LHC machine. We thank the technical and administrative staff at CERN and other CMS institutes, and acknowledge support from: FMSR (Austria); FNRS and FWO (Belgium); CNPq, CAPES, FAPERJ, and FAPESP (Brazil); MES (Bulgaria); CERN; CAS, MoST, and NSFC (China); COLCIENCIAS (Colombia); MSES (Croatia); RPF (Cyprus); Academy of Sciences and NICPB (Estonia); Academy of Finland, MEC, and HIP (Finland); CEA and CNRS/IN2P3 (France); BMBF, DFG, and HGF (Germany); GSRT (Greece); OTKA and NKTH (Hungary); DAE and DST (India); IPM (Iran); SFI (Ireland); INFN (Italy); NRF and WCU (Korea); LAS (Lithuania); CINVESTAV, CONACYT, SEP, and UASLP-FAI (Mexico); PAEC (Pakistan); SCSR (Poland); FCT (Portugal); JINR (Armenia, Belarus, Georgia, Ukraine, Uzbekistan); MST and MAE (Russia); MSTD (Serbia); MICINN and CPAN (Spain); Swiss Funding Agencies (Switzerland); NSC (Taipei); TUBITAK and TAEK (Turkey); STFC (United Kingdom); DOE and NSF (USA).
Individuals have received support from the Marie-Curie programme and the European Research Council (European Union); the Leventis Foundation; the A. P. Sloan Foundation; the Alexander von Humboldt Foundation; the Associazione per lo Sviluppo Scientifico e Tecnologico del Piemonte (Italy); the Belgian Federal Science Policy Office; the Fonds pour la Formation \`a la Recherche dans l'Industrie et dans l'Agriculture (FRIA-Belgium); and the Agentschap voor Innovatie door Wetenschap en Technologie (IWT-Belgium).

%% file: QCD-10-008-authorlist.tex
\textbf{Yerevan Physics Institute,  Yerevan,  Armenia}\\*[0pt]
S.~Chatrchyan, V.~Khachatryan, A.M.~Sirunyan, A.~Tumasyan
\vskip\cmsinstskip
\textbf{Institut f\"{u}r Hochenergiephysik der OeAW,  Wien,  Austria}\\*[0pt]
W.~Adam, T.~Bergauer, M.~Dragicevic, J.~Er\"{o}, C.~Fabjan, M.~Friedl, R.~Fr\"{u}hwirth, V.M.~Ghete, J.~Hammer\cmsAuthorMark{1}, S.~H\"{a}nsel, M.~Hoch, N.~H\"{o}rmann, J.~Hrubec, M.~Jeitler, W.~Kiesenhofer, M.~Krammer, D.~Liko, I.~Mikulec, M.~Pernicka, H.~Rohringer, R.~Sch\"{o}fbeck, J.~Strauss, A.~Taurok, F.~Teischinger, P.~Wagner, W.~Waltenberger, G.~Walzel, E.~Widl, C.-E.~Wulz
% ...
\vskip\cmsinstskip
% ...
1:~~Also at CERN, European Organization for Nuclear Research, Geneva, Switzerland\\
% ...
